\documentclass[preprint,showpacs,preprintnumbers,amsmath,amssymb,amssymb]{revtex4}

\usepackage{graphicx}
\usepackage{dcolumn}
\usepackage{bm}

\begin{document}

\hyphenpenalty=5000
\tolerance=1000
\makeatletter


\title{Neutral 3-3-1 Higgs Boson at LHC}   
\author{J.\ E.\ Cieza Montalvo$^1$}   
\affiliation{$^1$Instituto de F\'{\i}sica, Universidade do Estado do Rio de Janeiro, Rua S\~ao Francisco Xavier 524, 20559-900 Rio de Janeiro, RJ, Brazil}
\author{R. J. Gil Ram\'{i}rez, G. H. Ram\'{i}rez Ulloa, A. I. Rivasplata Mendoza$^2$}   
\affiliation{$^2$Universidad Nacional de Trujillo, Departamento de F\'{i}sica, Av. Juan Pablo II S/N; Ciudad Universitaria}  
\author{M. D. Tonasse$^{3}$\footnote{Permanent address: Universidade Estadual Paulista, {\it Campus} Experimental de Registro, Rua Nelson Brihi Badur 430, 11900-000 Registro, SP, Brazil}}   
\affiliation{$^3$Instituto de F\'\i sica Te\'orica, Universidade Estadual Paulista, \\  Rua Dr. Bento Teobaldo Ferraz 271, 01140-070 S\~ao Paulo, SP, Brazil}   
\date{\today}
   
\pacs{\\
11.15.Ex: Spontaneous breaking of gauge symmetries,\\
12.60.Fr: Extensions of electroweak Higgs sector,\\
14.80.Cp: Non-standard-model Higgs bosons.}
\keywords{Neutral Higgs, LHC, 331 model, branching ratio}
\begin{abstract}
We present an analysis of production and signature of neutral Higgs bosons on the version of the 3-3-1 model containing heavy leptons at the Large Hadron Collider (LHC). The production rate is found to be significant and the signal is clear, showing that these scalars can be detected in this accelerator. We also studied the possibility to identify them using their respective branching ratios. Cross section are given for two collider energies, $\sqrt{s} =$ 8 TeV and 14 TeV. Event rates and significances are discussed for two possible values of integrated luminosity, 10 fb$^{-1}$ and 300 fb$^{-1}$.

\end{abstract}
\maketitle

\section{INTRODUCTION \label{introd}}

The way to the understanding of the symmetry breaking mainly go 
through the scalars, although there are many other models that not contain elementary scalar fields, such as Nambu-Jona-Lasinio mechanism, technicolour theories, the strongly interacting gauge systems  \cite{nambu}. These  scalars protect the renormalizability of the theory by moderating the cross section growth. But so far, despite many experimental and theoretical efforts in order to understanding the scalar sector, the Higgs mechanism remains still unintelligible.  Nowadays, the major goal of the experimentalists in particle physics at the Large Hadron Collider (LHC), is to unravel the nature of electroweak symmetry breaking. The Standard Model (SM) is the prototype of a gauge theory with spontaneous symmetry breaking. This had great success in explaining the most of the experimental data.  However, recent results from neutrino osclillation experiments makes clear that the SM is not complete, then the neutrino oscillation implies that at least two neutrino flavors are massive. Moreover, there are others crucial problems in particle physics that not get response in SM. For instance, it offers not solution to the dark matter problem, dark energy and the asymmetry of matter-antimatter in the Universe. 
Therefore, there is a consensus among the particle physicists that the SM must be extended. \par 

In the SM appear only one elementary scalar,  which arises through the breaking of electroweak symmetry and this is the Higgs boson. The Higgs Boson is an important prediction of several quantum field theories and is so crucial to our understanding of the Universe. The Higgs boson is the one missing piece, is a critical ingredient to complete our understanding of the SM. Different types of Higgs bosons, if they exist, may lead us into new realms of physics beyond the SM. So, the observation of any kind of Higgs particle must be an important step forward in the understanding of physics in the electroweak sector or beyond the SM. \par 

Since the SM leaves many questions open, there are several  extensions. For example, if the Grand Unified Theory (GUT) contains the SM at high energies, then the Higgs bosons associated with GUT symmetry breaking must have masses of order $M_{X} \sim {\cal O} (10^{15})$
GeV. Supersymmetry \cite{supers} provides a solution to hierarchy problem through the cancellation of the quadratic divergences via fermionic and bosonic loops contributions \cite{cancell}. Moreover, the Minimal Supersymmetric extension of the SM (MSSM) can be derived as an effective theory of supersymmetric Grand Unified Theories \cite{sgut}. \par 

Among these extensions of the SM there are also other class of models based on SU(3)$_C \otimes$SU(3)$_L \otimes$U(1)$_N$ gauge symmetry (3-3-1 model) \cite{PT93a, PP92, FR92}, where the anomaly cancellation mechanisms occur when the three basic fermion families are considered and not family by family as in the SM. This mechanism is peculiar because it requires that the number of families is an integer multiple of the number of colors. This feature combined together  with the asymptotic freedom, which is a property of quantum chromodynamics (QCD), requires that the number of colors is less than five, and therefore as a consequence of this, the number of family of fermions must be exactly equal to three. Moreover, according to these models, the Weinberg angle is restricted to the value $s_W^2 = \sin^2\theta_W <1/4$ in the version of heavy-leptons \cite {PT93a}, but when evolves to higher values, it shows that the model loses its perturbative character when it reaches to mass scale of about 4 TeV \cite{DI05}. Then, the 3-3-1 model is one of the most interesting extensions of the SM and is phenomenologically well motivated to be probed at the LHC and other accelerators. \par 

In this work we study the production and signatures of two neutral Higgs bosons, predicted by the 3-3-1 model, which incorporates the charged heavy leptons \cite{PT93a, TO96}. One of these neutral Higgs is the standard one. We can show that the neutral Higgs boson signatures can be significant at LHC. Clear signal of these new particles can be obtained by studying the different decay modes. With respect to both mechanisms, that is the Drell-Yan and gluon-gluon fusion, we consider the $Z$, $Z^\prime$, $H_1^0$ and $H_2^0$ as propagators. Therefore in Sec. II we present the relevant features of the model. In Sec. III we compute the total cross sections of the process and in Sec. IV we summarize our results and conclusions.


\section{Relevant Features of the Model \label{sec2}}

We are working here with the version of the 3-3-1 model that contains heavy leptons \cite{PT93a}. The model is based on the semi simple symmetry group SU(3)$_C$$\otimes$SU(2)$_L$$\otimes$U(1)$_N$. The electric charge operator is given by

\begin{equation}
\frac{Q}{e} = \left(T_3 - \sqrt{3} \ T_8\right) + N,
\label{op}\end{equation}
where $T_3$ and $T_8$ are the generators of SU(3) and $e$ is the elementary electric charge. So, we can build three triplets of quarks of SU(3)$_L$ such that 
\begin{equation}
Q_{1L} = \left(\begin{matrix} u^\prime_1 \\ d^\prime_1 \\ J_1\end{matrix}\right)_L \sim \left({\bf 3}, \frac{2}{3}\right), \qquad Q_{\alpha L} = \left(\begin{matrix} J^\prime_\alpha \\ u^\prime_\alpha \\ d^\prime_\alpha\end{matrix}\right)_L \sim \left({\bf 3}^*, -\frac{1}{3}\right).
\label{quarks}\end{equation}
where the new quark $J_1$ carries $5/3$ units of electric charge while $J_\alpha$ $\left(\alpha = 2, 3\right)$ carry $-4/3$ each. We must also introduce the right-handed fermionic fields $U_R \sim \left({\bf 1}, 2/3 \right)$, $D_R \sim \left({\bf 1}, -1/3 \right)$, $J_{1R} \sim \left({\bf 1}, 5/3 \right)$ and $J^\prime_{\alpha R} \sim \left({\bf 1}, -4/3 \right)$. We have defined $U = \left(\begin{matrix}u^\prime & c^\prime & t^\prime\end{matrix}\right)$ and $D = \left(\begin{matrix}d^\prime & s^\prime & b^\prime\end{matrix}\right)$. \par

The spontaneous symmetry breaking is accomplished {\it via} three SU(3) scalar triplets, which are,
 
\begin{equation}
\eta = \left(\begin{matrix} \eta^0 \\ \eta^-_1 \\ \eta^+_2\end{matrix}\right) \sim \left({\bf 3}, 0\right), \quad \rho = \left(\begin{matrix} \rho^+ \\ \rho^0 \\ \rho^{++}\end{matrix}\right) \sim \left({\bf 3}, 1\right), \quad \chi = \left(\begin{matrix} \chi^- \\ \chi^{--} \\ \chi^0\end{matrix}\right) \sim \left({\bf 3}, -1\right).
\label{rc}\end{equation}

For sake of simplicity, we will assume here that the model respects the $B + L$ symmetry, where $B$ is the baryon number and $L$ is the lepton number. Then, the more general renormalizable Higgs potential is given by
\begin{eqnarray}
V\left(\eta, \rho, \chi\right) & = & \mu_1^2\eta^\dagger\eta + \mu_2^2\rho^\dagger\rho + \mu_3^2\chi^\dagger\chi +  \lambda_1\left(\eta^\dagger\eta\right)^2 + \lambda_2\left(\rho^\dagger\rho\right)^2 + \lambda_3\left(\chi^\dagger\chi\right)^2 + \cr 
&& + \eta^\dagger\eta\left[\lambda_4\left(\rho^\dagger\rho\right) + \lambda_5\left(\chi^\dagger\chi\right)\right] + \lambda_6\left(\rho^\dagger\rho\right)\left(\chi^\dagger\chi\right) + \lambda_7\left(\rho^\dagger\eta\right)\left(\eta^\dagger\rho\right) + \cr 
&& + \lambda_8\left(\chi^\dagger\eta\right)\left(\eta^\dagger\chi\right) + \lambda_9\left(\rho^\dagger\chi\right)\left(\chi^\dagger\rho\right) + \frac{1}{2}\left(f\varepsilon^{ijk}\eta_i\rho_j\chi_k + {\mbox{c. H.}}\right),
\label{pot}\end{eqnarray}
where $\mu_i$ $\left(i = 1, 2, 3\right)$ and $f$ are constants with mass dimension and $\lambda_j$ $\left(j = 1, \ldots, 9\right)$ are dimensionless constants \cite{TO96}. The potential (\ref{pot}) is bounded from below when the neutral Higgs fields develops the vacuum  expectation values (VEVs) $\langle\eta^0\rangle = v_\eta$, $\langle\rho^0\rangle = v_\rho$ and $\langle\chi^0\rangle = v_\chi$, with $v_\eta^2 + v_\rho^2 = v_W^2 = 246^2$ GeV$^2$. The scalar $\chi^0$ is supposedly heavy and it is responsible for the   spontaneous symmetry breaking of SU(3)$_L$$\otimes$U(1)$_N$ to SU(2)$_L$$\otimes$U(1)$_Y$ of the standard model. Meanwhile, $\eta^0$ and $\rho^0$ are lightweight and are responsible for the breaking of SU(2)$_L$$\otimes$U(1)$_Y$ to U(1)$_Q$, of the electromagnetism. Therefore, it is reasonable to expect 
\begin{equation}
v_\chi \gg v_\eta, v_\rho.
\label{ap}\end{equation}
The potential (\ref{pot}) provides the masses of neutral Higgs as
\begin{subequations}\begin{align}
m^2_{H_1^0} & \approx 4\frac{\lambda_2 v_\rho^4 - \lambda_1 v_\eta^4}{v_\eta^2 - v_\rho^2}, & m^2_{H_2^0} &\approx \frac{v_W^2v_\chi^2}{2v_\eta v_\rho}, \\
m^2_{H_3^0} & \approx -\lambda_3 v_\chi, & m^2_h & = -\frac{fv_\chi}{v_\eta v_\rho}\left[v_W^2 + \left(\frac{v_\eta v_\rho}{v_\chi}\right)^2\right]
\end{align}\label{mas}\end{subequations}

with the corresponding eigenstates
\begin{equation}
\left(\begin{matrix} \xi_\eta \\ \xi_\rho\end{matrix}\right) \approx \left(\begin{matrix} c_w & s_w \\ s_w & c_w\end{matrix}\right)\left(\begin{matrix} H_1^0 \\ H_2^0 \end{matrix}\right), \quad \xi_\chi \approx H_3^0, \quad \zeta_\chi \approx ih,
\label{auto}\end{equation}
where the mixing parameters are $c_w = \cos w = v_\eta/\sqrt{v_\eta^2 + v_\rho^2}$ and $s_w = \sin w$ \cite{TO96}. In Eqs. (\ref{mas}) and (\ref{auto}) we used the approximation (\ref{ap}) and, for not to introduce the new mass scale in the model, we assume $f \approx -v_\chi$. We can then note that $H^0_3$ is a typical 3-3-1 Higgs boson. The scalar $H^0_1$ is one that can be identified with the SM Higgs, since its mass and eigenstate do not depend on $v_\chi$. \par
Now, we can write the Yukawa interactions for the ordinary quarks, i.e.
\begin{eqnarray}
{\cal L}^Y_{q} & = & \sum_\alpha\left[\overline{Q}_{1L}\left(G_{1\alpha}U^\prime_{\alpha R}\eta + \tilde{G}_{1\alpha}D^\prime_{\alpha R}\rho\right) + \sum_i\overline{Q}_{iL}\left(F_{i\alpha}U^\prime_{\alpha R}\rho^* + \tilde{F}_{i\alpha}D^\prime_{\alpha R}\eta^*\right)\right],
\label{Lq}
\end{eqnarray}
where $G_{ab}$ and $G^\prime_{ab}$ ($a$ and $b$ are generation indexes) are coupling constants. \par
The interaction eigenstates (\ref{quarks}) and their right-handed counterparts can rotate about their respective physical eigenstates as 
\begin{subequations}\begin{eqnarray}
&& U^\prime_{aL\left(R\right)} = {\cal U}_{ab}^{L\left(R\right)}U_{bL\left(R\right)}, \\
&& D^\prime_{aL\left(R\right)} = {\cal D}_{ab}^{L\left(R\right)}U_{bL\left(R\right)}, \quad J^\prime_{aL\left(R\right)} = {\cal J}_{ab}^{L\left(R\right)}J_{bL\left(R\right)},
\end{eqnarray}\end{subequations}

Since the cross sections involving the sum over the flavors and rotation matrices are unitary, then the mixing parameters have no major effect on the calculations. In terms of physical fields the Yukawa Lagrangian for the neutral Higgs can be written as

\begin{eqnarray}
-{\cal L}_Q & = & \frac{1}{2}\left\{\overline{U}\left(1 + \gamma_5\right)\left[1 + \left[\frac{s_w}{v_\rho} + \left(\frac{c_w}{v_\eta} + \frac{s_w}{v_\rho}\right){\cal V}^U\right]H_1^0 + \right.\right. \cr
&& \left.\left. + \left[\frac{-c_w}{v_\rho} + \left(\frac{s_w}{v_\eta} - \frac{c_w}{v_\rho}\right){\cal V}^U\right]H_2^0\right]M^UU + \right. \cr
&& \left. + \overline{D}\left(1 + \gamma_5\right)\left[1 + \left[\frac{c_w}{v_\eta} + \left(\frac{s_w}{v_\rho} - \frac{c_w}{v_\eta}\right){\cal V}^D\right]H_1^0 + \right.\right. \cr
&& \left.\left. + \left[\frac{s_w}{v_\eta} - \left(\frac{c_w}{v_\rho} + \frac{s_w}{v_\eta}\right){\cal V}^D\right]H_2^0\right] M^DD\right\} + {\mbox{H. c.}},
\end{eqnarray}\label{Yuk}
where $V^U_LV^D_L = V_{\rm CKM}$ is the Cabibbo-Kobayashi-Maskawa matrix, ${\cal V}^U$ and ${\cal V}^D$ are arbitrary mixing matrices and $M^U = {\rm diag}\left(\begin{matrix} m_u & m_c & m_t\end{matrix}\right)$ and $M^D = {\rm diag}\left(\begin{matrix} m_d & m_s & m_b\end{matrix}\right)$ are matrices which carrying the masses of the quarks. \par

In the gauge sector, beyond the standard particles $\gamma$, $Z$, and $W^\pm$ the model predicts: one neutral $\left(Z^\prime\right)$, two single-charged $\left(V^\pm\right)$, and two doubly-charged $\left(U^{\pm\pm}\right)$ gauge bosons. The gauge interactions with  Higgs  bosons are given by 
\begin{equation}
{\cal L}_{GH} = \sum_\varphi\left({\cal D_\mu\varphi}\right)^\dagger\left({\cal D_\mu\varphi}\right),
\label{deri}\end{equation}
where the covariant derivatives are
\begin{equation}
{\cal D}_\mu\varphi_i = \partial_\mu\varphi_i - ig\left({W}_\mu.\frac{{ T}}{2}\right)^j_i\varphi_j - ig^\prime N_\varphi\varphi_iB_\mu,
\end{equation}
where $\varphi = \eta$, $\rho$, $\chi$ $\left(N_\eta = 0, N_\rho = 1, N_\chi = -1\right)$ are the Higgs triplets, ${W}_\mu$ and $B_\mu$ are the SU(2) and U(1) field tensors, $g$ and $g^\prime$ are the U(1) and SU(2) coupling constants, respectively. Diagonalization of the Lagrangean (\ref{deri}), after symmetry 
breaking, gives masses for the neutral weak gauge bosons, i.e.,

\begin{equation}
m_Z \approx \frac{\vert e\vert}{2s_Wc_W}v_W,     \qquad m_{Z^\prime}^2 \approx \frac{1}{3\left(1 - 4s_W^2\right)}\left(\frac{\vert e\vert c_Wv_\chi}{s_W}\right)^2,
\label{masszz}\end{equation}

where $s_W = \sin{\theta_W}$, with $\theta_W$ being the Weinberg angle, and $c_W^2 = 1 - s_W^2$. Then the eigenstates are
\begin{subequations}\begin{eqnarray}
W^3_\mu & \approx & s_WA_\mu - c_WZ_\mu \\
W^8_\mu & \approx & -\sqrt{3}s_W\left(A_\mu - \frac{s_W}{c_W}Z_\mu\right) + \frac{\sqrt{1 - 4s_W^2}}{c_W}Z^\prime_\mu \\
B_\mu & \approx & \frac{s_W}{\sqrt{1 - 4s_W^2}}A_\mu + \frac{s_W}{c_W}\left(Z_\mu + \sqrt{3}Z^\prime_\mu\right).
\end{eqnarray}\label{eig}\end{subequations}
In Eqs. (\ref{masszz}) and (\ref{eig}) we have used the approximation (\ref{ap}). Finally the weak neutral current in the sector of u and d quarks reads 

\begin{subequations}\begin{eqnarray}
-{\cal L}_Z & = & \frac{\left|e\right|}{2s_Wc_W}\overline{q}\gamma^\mu\left[v\left(q\right) + a\left(q\right)\gamma_5\right]qZ_\mu \\
-{\cal L}_{Z^\prime} & = & \frac{\left|e\right|}{2s_Wc_W}\overline{q}\gamma^\mu\left[v^\prime\left(q\right) + a^\prime\left(q\right)\gamma_5\right]qZ^\prime_\mu
\end{eqnarray}\label{ZZl}\end{subequations}
whose coefficients are
\begin{subequations}\begin{align}
v\left(u\right) & = 1 - \frac{s_W^2}{8}, & a\left(u\right) & = -a\left(d\right) = -1, & v\left(d\right) & = -1 + \frac{4}{3}s_W^2, \\
v^\prime\left(u\right) & = \sqrt{1 + 4s_W^2}, & a^\prime\left(u\right) & = \sqrt{\frac{1 - 4s_W^2}{3}},  & v^\prime\left(d\right) & = \frac{2s_W^2 - 1}{\sqrt{3}}, \\
a^\prime\left(d\right) & = -v^\prime\left(d\right).
\end{align}\label{coefq}\end{subequations}

In this work we study the production of a neutral Higgs bosons at $pp$ colliders. With respect to both mechanisms, that is the Drell-Yan and gluon-gluon fusion, we consider the $Z$, $Z^\prime$, $H_1^0$ and $H_2^0$ as propagators.

\section{CROSS SECTION PRODUCTION}   

The mechanisms for the production of neutral Higgs particles in $pp$ collisions occurs in association with the boson  Z, $Z^{\prime}$, $H_{1}^{0}$ and $H_{2}^{0}$, see Fig. 1 and Fig. 2. Contrarily to the standard model, where the gluon-gluon fusion dominates over Drell-Yan when the Higgs boson is heavier than $100$ GeV \cite{barger}, in 3-3-1 Model does  not occur, the mechanism of Drell-Yan dominates over gluon-gluon fusion at leading order (LO) for $H_{1}^{0}$ and $H_{2}^{0}$ production at $\sqrt{s}=8(14)$ TeV. The process $pp \to  H_{i}^{0} Z$ $(i = 1, 2)$ takes place in the $s$ channel. So using the interaction Lagrangian \cite{PT93a, cieton2}, we obtain the differential cross section in the first place for Drell-Yan for $H_{1}^{0}$ and $H_{2}^{0}$

\begin{eqnarray}  
\frac{\hat{d \sigma_{H_{1}^{0}}}}{d \Omega} & = & \frac{1}{64 \pi^{2} \hat{s}} \left(  \overline{\left|A_{Z}\right|^{2}} + \overline{\left|A_{Z^{\prime}}\right|^{2}} + \overline{\left|A_{H_{1}^{0}}\right|^{2}} +\overline{\left|A_{H_{2}^{0} }\right|^{2}} +2 {\it Re} \overline{A_{Z}^{*} \ {A_{Z^{\prime}}}}  + 2 {\it Re} \overline{A_{H_{1}^{0}}^{*} \ {A_{H_{2}^{0}}}} \right), \nonumber  \\
\end{eqnarray}   

\vskip -1cm

\begin{eqnarray}  
\frac{\hat{d \sigma_{H_{2}^{0}}}}{d \Omega} & = & \frac{1}{64 \pi^{2} \hat{s}} \left( \overline{\left|A_{Z^{\prime}}\right|^{2}} + \overline{\left|A_{H_{2}^{0}}\right|^{2}} + \overline{\left|A_{H_{1}^{0} }\right|^{2}} + 2 {\it Re} \overline{A_{H_{1}^{0}}^{*} \ {A_{H_{2}^{0}}}} \right). \nonumber  \\  \ ,
\end{eqnarray} 
where the $A_{i}$ is the matrix element of each particle. \par 

In the cross section for $H_{i}^{0}$ production, where $i=1,2$, the interference term between the $Z(Z^{\prime})$ and $H_{i}^{0}$ should be absent because it gives an imaginary value, and in the cross section for $H_{2}^{0}$ production   the term involving the boson $Z$ is absent, because there is no coupling between the $Z$ and $H_{2}^{0} \ Z$, then we write  separately the differential cross section for $H_{1}^{0}$ and $H_{2}^{0}$ production

\begin{eqnarray} 
\frac{d \hat{\sigma}_{H_{1}^{0}}}{d\cos \theta} & = &\frac{\beta_{H_{1}^{0}} \ g^{2}}{192 \pi \cos^{2} {\theta_{W}} s} \Biggl \{\frac{m_{W}^{2} \ g^{2}}{4 cos^{4} {\theta_{W}} (s- m_{Z}^{2} + i m_Z \Gamma_Z)^{2}} \left ((m_{Z}^{2}+ \frac{tu}{m_{Z}^{2}}- t- u + s)(g_{{V}^{q}}^{2}+ g_{{A}^{q}}^{2}) \right )  \nonumber \\
&&  + \frac{g^4 \ \Lambda_{ZZ'H_{1}^{0}}^{2}}{4(s- m_{Z'}^{2}+ im_{Z'} \Gamma_{Z'})^{2}}
\left ((m_{Z}^{2}+ \frac{tu}{m_{Z}^{2}}- t- u + s)(g_{{V'}^{q}}^{2}+ g_{{A'}^{q}}^{2}) \right )   \nonumber  \\ 
&&  +  \frac{g^3 \ m_{W}   \ \Lambda_{ZZ'H_{1}^{0}}}{2 \ cos^{2} {\theta_{W}} (s- m_{Z}^{2}+ im_{Z} \Gamma_{Z}) (s- m_{Z'}^{2}+ im_{Z'} \Gamma_{Z'})}  ((m_{Z}^{2}+ \frac{tu}{m_{Z}^{2}}- t- u + s)   \nonumber \\ 
&&(g_{V}^{q}  g_{V'}^{q}+ g_{A}^{q} g_{A'}^{q}) ) + \left( \frac{m_{q}^{2} \ (v_{\rho}^{2}- v_{\eta}^{2})^2}{32 \ v_{W}^{6}} |\chi^{(1)}(\hat{s})|^{2} 
 +  \frac{v_{\eta}^{2} \ v_{\rho}^{2}}{2 \ v_{W}^{6}}  \left (m_{u} \frac{v_{\eta}}{v_{\rho}}- m_{d} \frac{v_{\rho}}{v_{\eta}} \right )^{2} |\chi^{(2)}(\hat{s})|^{2}  \right.   \nonumber  \\
&&\left.  + \frac{m_{q}  \left (m_{u} \frac{v_{\eta}}{v_{\rho}}- m_{d} \frac{v_{\rho}}{v_{\eta}} \right ) \ v_{\eta} v_{\rho} (v_{\rho}^{2}-v_{\eta}^{2})}{4 \ v_{W}^{6}} |\chi^{(1)}(\hat{s})|  |\chi^{(2)}(\hat{s})| \right)  \nonumber \\   
&& \left(\frac{\hat{s}}{m_{Z}^{2}} (\hat{s}^{2}-2 m_{Z}^{2} \ \hat{s}+  m_{Z}^{4}) - \frac{m_{q}^{2}}{m_{Z}^{2}} (2 \hat{s}^{2} - 4 \hat{s} \ m_{Z}^{2} + 2  m_{Z}^{4} )  \right. +  \nonumber \\
&&\left. -\frac{m_{H_{1}^{0}}^{2}}{m_{Z}^{2}} (2 \hat{s}^{2} + 2 \hat{s} m_{q}^{2}- 4 m_{q}^{2} \hat{s} - 4 m_{q}^{2} m_{Z}^{2} ) + \frac{m_{H_{1}^{0}}^{4} \hat{s}}{m_{Z}^{2}}- \frac{2 m_{H_{1}^{0}}^{4} m_{q}^{2}}{m_{Z}^{2}}     \  \right)  \Biggr \}  , 
\label{DZZ'H}
\end{eqnarray} 

\begin{eqnarray} 
\frac{d \hat{\sigma}_{H_{2}^{0}}}{d\cos \theta} & = &\frac{\beta_{H_{2}^{0}} \ g^{2}}{192 \pi \cos^{2} {\theta_{W}} s} \Biggl \{ \frac{g^4 \ \Lambda_{ ZZ'H_{2}^{0}}^{2}}{4(s- m_{Z'}^{2}+ im_{Z'} \Gamma_{Z'})^{2}}
\left ((m_{Z}^{2}+ \frac{tu}{m_{Z}^{2}}- t- u + s)(g_{{V'}^{q}}^{2}+ g_{{A'}^{q}}^{2}) \right )   \nonumber  \\ 
&& + \left( \frac{v_{\eta}^{2} \ v_{\rho}^{2} m_{q}^{2} }{2 \ v_{W}^{6}}  |\chi^{(1)}(\hat{s})|^{2} + \frac{\left (m_{u} \frac{v_{\eta}}{v_{\rho}}- m_{d} \frac{v_{\rho}}{v_{\eta}} \right )^{2} (v_{\rho}^{2}- v_{\eta}^{2})^2}{32 \ v_{W}^{6}} |\chi^{(2)}(\hat{s})|^{2} \right.  \nonumber  \\
&&\left.  + \frac{m_{q}  \left (m_{u} \frac{v_{\eta}}{v_{\rho}}- m_{d} \frac{v_{\rho}}{v_{\eta}} \right ) \ v_{\eta} v_{\rho} (v_{\rho}^{2}-v_{\eta}^{2})}{4 \ v_{W}^{6}} |\chi^{(1)}(\hat{s})|  |\chi^{(2)}(\hat{s})| \right)  \nonumber \\   
&& \left(\frac{\hat{s}}{m_{Z}^{2}} (\hat{s}^{2}-2 m_{Z}^{2} \ \hat{s}+  m_{Z}^{4}) - \frac{m_{q}^{2}}{m_{Z}^{2}} (2 \hat{s}^{2} - 4 \hat{s} \ m_{Z}^{2} + 2  m_{Z}^{4} )  \right. +  \nonumber \\
&&\left. -\frac{m_{H_{2}^{0}}^{2}}{m_{Z}^{2}} (2 \hat{s}^{2} + 2 \hat{s} m_{q}^{2}- 4 m_{q}^{2} \hat{s} - 4 m_{q}^{2} m_{Z}^{2} ) + \frac{m_{H_{2}^{0}}^{4} \hat{s}}{m_{Z}^{2}}- \frac{2 m_{H_{2}^{0}}^{4} m_{q}^{2}}{m_{Z}^{2}}     \  \right)  \Biggr \}  , 
\label{DZZ'H2}
\end{eqnarray} 
here g is the coupling constant of the weak interaction, the $\beta_{H_{i}^{0}}$ ($i=1,2$) is the Higgs velocity in the c.m. of the subprocess which is equal to
\[
\beta_{H_{i}^{0}} = \frac{ \left [\left( 1- \frac{(m_{Z}+ m_{H_{i}^{0}})^{2}}{\hat{s}} \right) \left(1- \frac{(m_{Z}- m_{H_{i}^{0}})^{2}}{\hat{s}} \right) \right ]^{1/2}}{1-\frac{m_{Z}^{2}-m_{H_{i}^{0}}^{2}}{\hat{s}}}  \ \ ,
\]

and we have also defined

\[
\chi^{i}(\hat{s}) = \frac{1}{\hat{s}-m_{H_{i}^{0}}^{2} + i m_{H_{i}^{0}} \Gamma_{H_{i}^{0}}}  \ ,
\]
with $\Gamma_{H_{i}^{0}}$ being the Higgs boson total width, the primes $\left(^\prime\right)$ are for the case when we take a $Z'$ boson, $\Gamma_{Z}$ and $\Gamma_{Z'}$ \cite{cieton1,cieton2}, are the total width of the $Z$ and $Z'$ boson, $g_{V, A}^{q}$ are the standard quark coupling constants, $m_{q}$, were $q=u,d$ are the masses of the quark, $g_{V', A'}^{q}$ are the 3-3-1 quark coupling constants, $\sqrt{\hat{s}}$ is the center of mass energy of the  $q \bar{q}$ system, $g= \sqrt{4 \ \pi \ \alpha}/\sin \theta_{W}$ and $\alpha$ is the fine structure constant, which we take equal to $\alpha=1/128$. For the $Z^\prime$ boson we take  $M_{Z^\prime} = \left(0.5 - 3\right)$ TeV, since $M_{Z^\prime}$ is proportional to the VEV $v_\chi$ \cite{PP92,FR92}. For the standard model parameters we assume PDG values, {\it i. e.}, $M_Z = 91.19$ GeV, $\sin^2{\theta_W} = 0.2315$, and $M_W = 80.33$ GeV 
\cite{Nea10}, t and u are the kinematic invariants. We have also defined the $\Lambda_{ZZH_{i}^{0}} (\Lambda_{ZZ'H_{i}^{0}})$ as the coupling constants of the $Z(Z^{\prime})$ boson to Z boson and Higgs $H_{i}^{0}$ where i stands for $H_{1}^{0}, H_{2}^{0}$, the $\Lambda_{H_{i}^{0}H_{i}^{0}Z}$ are the couplings constants of the  $H_{1}^{0}$ boson to $H_{1}^{0}$ and Z boson, of the  $H_{2}^{0}$ boson to $H_{2}^{0}$ and Z boson and of the  $H_{1}^{0}$ boson to $H_{2}^{0}$ and Z boson, these three coupling constants should be multiplied by $p^{\mu}-q^{\mu}$ to get a $\Lambda_{ZZH_{i}^{0}}^{\mu} = \Lambda_{ZZH_{i}^{0}} (p^{\mu}-q^{\mu})$ with $p$  and $q$ being the momentum four-vectors of the $H_{i}$ and $Z$ boson where $i=1,2$, and the $\Lambda_{q \bar{q} H_{i}^{0}}$ are the coupling constants of the $H_{1}^{0}(H_{2}^{0})$ to $q \bar{q}$, the $v\left(q\right)$ $a\left(q\right)$ $v^\prime\left(q\right)$ and $a^\prime\left(q\right)$ are given in \cite{cieton1}. It can be noticed then that the coupling $H_1^0Z_\mu Z_\nu$ is the same as the SM, as one would expect, while $H_2^0Z_\mu Z_\nu$ does not exist. We remark still that in 3-3-1 model, the states $H_1^0$ and $H_2^0$ are mixed. \par 

\[
t  = m_{q_{i}}^{2}+ m_{Z}^{2} - \frac{s}{2} \Biggl \{ \left(1+ \frac{m_{Z}^{2}- m_{H}^{2}}{s}\right)- \cos \theta  \left [\left( 1- \frac{(m_{Z}+ m_{H})^{2}}{s} \right) \left(1- \frac{(m_{Z}- m_{H})^{2}}{s} \right) \right ]^{1/2}\Biggr \}, 
\]

\[  
u  = m_{q_{i}}^{2}+ m_{H}^{2} - \frac{s}{2} \Biggl \{ \left(1- \frac{m_{Z}^{2}- m_{H}^{2}}{s}\right)+ \cos \theta  \left [\left( 1- \frac{(m_{Z}+ m_{H})^{2}}{s} \right) \left(1- \frac{(m_{Z}- m_{H})^{2}}{s} \right) \right ]^{1/2}\Biggr \}, 
\]

\begin{subequations}\begin{eqnarray}
\left(\Lambda_{q\bar{q}Z}\right)_\mu & \approx & i\frac{\vert e\vert}{2s_Wc_W}\gamma_\mu\left[v\left(q\right) + a\left(q\right)\gamma_5\right], \\
\left(\Lambda_{q\bar{q}Z^\prime}\right)_\mu & \approx & i\frac{\vert e\vert}{2s_Wc_W}\gamma_\mu\left[v^\prime\left(q\right) + a^\prime\left(q\right)\gamma_5\right], \\
\Lambda_{q\bar{q}H_1^0} & \approx & -i\frac{m_q}{2v_W}\left(1 + \gamma_5\right), \\
\Lambda_{q\bar{q}H_2^0} & \approx & \frac{i}{2v_W}\left(-m_u\frac{v_\eta}{2v_\rho} + m_d\frac{v_\rho}{v_\eta}\right)\left(1 + \gamma_5\right) \\
\left(\Lambda_{ZZH_1^0}\right)_{\mu\nu} & \approx & -i\frac{g^2v_W}{2}\left(\frac{m_Z}{m_W}\right)^2g_{\mu\nu}, \\
\left(\Lambda_{ZZ^\prime H_1^0}\right)_{\mu\nu} & \approx & -\frac{g^2}{2\sqrt{3}v_W}\frac{m_Z}{m_W}\frac{\left(1 + 2s_W^2\right)v_\eta^2 - \left(1 - 4s_W^2\right)v_\rho^2}{c_W\sqrt{1 - 4s_W^2}}g_{\mu\nu}, \\
\left(\Lambda_{ZZ^\prime H_2^0}\right)_{\mu\nu} & \approx & \frac{g^2}{\sqrt{3}\left(1 - 4s_W^2\right)}\frac{v_\eta v_\rho}{v_W}g_{\mu\nu}, \\ 
\left(\Lambda_{H_1^0H_1^0Z}\right)_\mu & \approx & \frac{g}{2}\frac{m_Z}{m_W}\frac{\left(v_\rho^2 - v_\eta^2\right)}{v_W^2}\left(p - q\right)_\mu, \\ 
\left(\Lambda_{H_2^0H_2^0Z}\right)_\mu & \approx & -\frac{g}{2}\frac{m_Z}{m_W}\frac{\left(v_\rho^2 - v_\eta^2\right)}{v_W^2}\left(p - q\right)_\mu, \\ 
\left(\Lambda_{H_1^0H_2^0Z}\right)_\mu & \approx & -2g\frac{m_Z}{m_W}\frac{v_\rho v_\eta}{v_W^2}\left(p - q\right)_\mu,
\label{eigc}\end{eqnarray}\label{eigthen}\end{subequations}

where $\theta$ is the angle between the Higgs and the incident quark in the CM frame. \par 

The total cross section for the process $pp \rightarrow qq \rightarrow Z H_{i}$ is related to the subprocess $qq \rightarrow Z H_{i}$ total cross  section $\hat{\sigma}$, through   
\begin{equation}    
\sigma = \int_{\tau_{min}}^{1} \int_{\ln{\sqrt{\tau_{min}}}}^{-\ln{\sqrt{\tau_{min}}}} d \ \tau \ dy \   q\left(\sqrt{\tau}e^y, Q^2\right) q\left(\sqrt{\tau}e^{-y}, Q^2\right)   \hat{\sigma}\left(\tau, s\right),   
\end{equation}\noindent
where $\tau_{min} = (m_{Z}+ m_{H_{i}})^{2}/s (\tau =\hat{s}/s )$ and  $q\left(x,Q^2\right)$ is the 
quark structure function.\par 

Another form to produce a neutral Higgs is {\it via} the gluon-gluon  fusion, namely through the  reaction of the type $pp \rightarrow g g   \rightarrow Z H_{i}$. Since the final state is neutral, the $s$ channel involves  the exchange of the boson $Z$, $Z'$, $H_{1}^{0}$ and $H_{2}^{0}$. The exchange of a photon is not allowed by C conservation (Furry's theorem), which also indicates that only the axial-vector couplings of the bosons $Z$ nad $Z'$, contribute to this process . Therefore, the differential cross section for production of $H_{1}^{0}$ and $H_{2}^{0}$ we make separately in order to do explicit the different contributions: \par 

\begin{eqnarray}  
\frac{\hat{d \sigma_{H_{1}^{0}}}}{d \Omega} & = & \frac{1}{64 \pi^{2} \hat{s}} \left(  \overline{\left|A_{Z}\right|^{2}} + \overline{\left|A_{Z^{\prime}}\right|^{2}} + 
\overline{\left|A_{H_{1}^{0}}\right|^{2}} +
\overline{\left|A_{H_{2}^{0}}\right|^{2}}  +2 {\it Re} \overline{A_{Z}^{*} \ {A_{Z'}}} +2 {\it Re} \overline{A_{Z(Z^{\prime})} \ {A_{H_{1}^{0}}^{*}}}  \right.  \nonumber  \\
&& \left.  +2 {\it Re} \overline{A_{Z(Z^{\prime})} \ {A_{H_{2}^{0}}^{*}}}  + 2 {\it Re} \overline{A_{H_{1}^{0}}^{*} \ {A_{H_{2}^{0}}}} 
\right), 
\end{eqnarray}   

\vskip -0.5cm

\begin{eqnarray}  
\frac{\hat{d \sigma_{H_{2}^{0}}}}{d \Omega} & = & \frac{1}{64 \pi^{2} \hat{s}} \left( \overline{\left|A_{Z^{\prime}}\right|^{2}} + \overline{\left|A_{H_{1}^{0}}\right|^{2}} + 
\overline{\left|A_{H_{2}^{0}}\right|^{2}} + 
2 {\it Re} \overline{A_{Z^{\prime}} \ {A_{H_{1}^{0}}^{*}}}  + 
2 {\it Re} \overline{A_{Z^{\prime}} \ {A_{H_{2}^{0}}^{*}}}   \right. \nonumber  \\
&& \left.  + 2 {\it Re} \overline{A_{H_{1}^{0}}^{*} \ {A_{H_{2}^{0}}}} 
 \right). 
\end{eqnarray}
where the $A_{i}$ is the matrix element of each particle. \par 

It is important to emphasis that for production of $H_{1}^{0}$ the interference term between the $Z(Z^{\prime})$, which are antisymmetric in the gluon polarizations, and the $H_{1}^{0}$ diagrams, we only consider the antisymmetric term of $H_{1}^{0}$, because the other part is symmetric and therefore vanishes, and in the production of $H_{2}^{0}$ we take the interference between the $Z^{\prime}$ diagrams and the antisymmetric term of $H_{2}^{0}$, then we write explicity the $Z, Z^{\prime}, H_{1}^{0}$ and $H_{2}^{0}$ contributions to the elementary cross section  

\begin{eqnarray} 
\left (\frac{d \hat{\sigma}}{d\cos \theta} \right )_{pp \rightarrow Z H_{i}^{0}}^{Z(Z')} & = & \frac{g^{4}(g^{6}) \ \alpha_s^2  \ (\Lambda_{Z(Z^{\prime})ZH_{i}^{0}})^{2} \ \Delta}{8192 \ \pi^{3}  \hat{s} \  \cos^{2}_{W}  M_{Z(Z')}^4} \beta_{H_{i}^{0}} \left|
\sum_{q=u,d} T_3^q(q') \left( 1 + 2 \delta_q I_q  \right) \right|^2   ,
\end{eqnarray}

\begin{eqnarray} 
\left (\frac{d \hat{\sigma}}{d\cos \theta} \right )_{pp \rightarrow Z H_{1}^{0}}^{H_{1}^{0}} & = &   \frac{g^{2} \ \alpha_s^2 \  (v_{\rho}^{2}-  v_{\eta}^{2})^{2} \ \hat{s} \ \Omega \ \beta_{H_{i}^{0}}}{131072  \ \pi^{3}  v_{W}^{6}  \  \cos^{2}_{\theta_{W}}}  \ |\chi^{(1)}(\hat{s})|^{2}  \left|  
\sum_{q=u,d}  \left[ 2 \delta_q + \delta_q (4 \ \delta_q-1)  I_q  \right] \right|^2   + \nonumber   \\
&& + \frac{g^{2} \ \alpha_s^2 \ (v_{\rho}^{2} - v_{\eta}^{2})^{2}  \ \Omega \  \beta_{H_{i}^{0}}}{65536 \ \pi^{3}  \hat{s} \  v_{W}^{6}  \cos^{2}_{W}} \ |\chi^{(1)}(\hat{s})|^{2} \left|\sum_{q=u,d}  m_{q}^{2} \  I_q  \right|^2   ,
\end{eqnarray}

\begin{eqnarray} 
\left (\frac{d \hat{\sigma}}{d\cos \theta} \right )_{pp \rightarrow Z H_{1}^{0}}^{H_{2}^{0}} & = &   \frac{g^{2} \ \alpha_s^2 \  v_{\eta}^{2} \ v_{\rho}^{2} \ \hat{s} \ \Omega \ \beta_{H_{i}^{0}}}{8192 \pi^{3}  v_{W}^{6}  \  \cos^{2}_{\theta_{W}}}  \ |\chi^{(2)}(\hat{s})|^{2}  \left|
\sum_{q=u,d} \frac{\left(m_{u} \frac{v_{\eta}}{v_{\rho}}- m_{d} \frac{v_{\rho}}{v_{\eta}} \right)}{m_{q}} \left[ 2 \delta_q + \delta_q (4 \ \delta_q-1)  I_q  \right] \right|^2   + \nonumber   \\
&& + \frac{g^{2} \ \alpha_s^2 \ v_{\eta}^{2} \ v_{\rho}^{2} \ \Omega \  \beta_{H_{i}^{0}}}{4096 \ \pi^{3} \hat{s} \  v_{W}^{6} \ \cos^{2}_{W}} \ |\chi^{(2)}(\hat{s})|^{2} \left|\sum_{q=u,d}  m_{q} \left (m_{u} \frac{v_{\eta}}{v_{\rho}}- m_{d} \frac{v_{\rho}}{v_{\eta}} \right ) I_q  \right|^2      ,
\end{eqnarray}

\begin{eqnarray} 
\left (\frac{d \hat{\sigma}}{d\cos \theta} \right )_{pp \rightarrow Z H_{1}^{0}}^{Z-Z'} & = & \frac{g^{5} \ \alpha_s^2   \ \Lambda_{ZZ'H_{1}^{0}} \ \Delta}{4096 \ \pi^{3}  \hat{s} \  \cos^{2}_{\theta_{W}} m_{Z}^2  m_{Z'}^{2}} \beta_{H_{i}^{0}}   \nonumber   \\
&& {\it Re} \left[
\sum_{q=u,d} T_3^q \left( 1 + 2 \delta_q I_q^{*}  \right)   
 \sum_{q=u,d} T_3^{q'} \left( 1 + 2 \delta_{q'} I_{q'}  \right) \right]          ,
\end{eqnarray}

\begin{eqnarray} 
\left (\frac{d \hat{\sigma}}{d\cos \theta} \right )_{pp \rightarrow Z H_{1}^{0}}^{Z(Z^{\prime})-H_{1}^{0}} & = & - \frac{g^{3}(g^{4}) \ \alpha_s^2   \ \Lambda_{Z(Z^{\prime})ZH_{1}^{0}} \ (v_{\rho}^{2} - v_{\eta}^{2}) \ \Pi \ \beta_{H_{i}^{0}}}{4096 \ \pi^{3} \ \hat{s} \ \cos^{2}_{\theta_{W}} v_{W}^{3}} {\it Re} \left[ \frac{\chi^{(1)}(\hat{s})}{(s- m_{Z(Z^{\prime})}^{2}+ im_{Z(Z^{\prime})} \Gamma_{Z(Z^{\prime})})}    \right.  \nonumber  \\
&& \left. \sum_{q=u,d}  m_{q}^{2} \ T_3^q \left( 1 + 2 \delta_q I_q  \right) \ \ \sum_{q=u,d}  I_q^{*} \right]          ,
\end{eqnarray}

\begin{eqnarray} 
\left (\frac{d \hat{\sigma}}{d\cos \theta} \right )_{pp \rightarrow Z H_{1}^{0}(H_{2}^{0})}^{H_{1}^{0}-H_{2}^{0}} & = & \mp \frac{g^{2} \ \alpha_s^2 \ (v_{\rho}^{2} - v_{\eta}^{2}) v_{\rho} v_{\eta} \ \Omega \ \beta_{H_{i}^{0}}}{8192 \ \pi^{3} \ \hat{s} \  \cos^{2}_{\theta_{W}} v_{W}^{6}} {\it Re}  \  \chi^{(1)}(\hat{s}) \ \chi^{(2)}(\hat{s})     \nonumber  \\
&& \sum_{q=u,d}  m_{q}^{3} \left (m_{u} \frac{v_{\eta}}{v_{\rho}}- m_{d} \frac{v_{\rho}}{v_{\eta}} \right )   I_q  \ \sum_{q=u,d}  I_q^{*}   \nonumber \mp  \frac{g^{2} \ \alpha_s^2 \ (v_{\rho}^{2} - v_{\eta}^{2}) v_{\rho} v_{\eta} \ s \ \Omega \ \beta_{H_{i}^{0}}}{16384\ \pi^{3}  \  \cos^{2}_{\theta_{W}} v_{W}^{6}}     \nonumber  \\
&&  {\it Re}  \  \chi^{(1)}(\hat{s}) \ \chi^{(2)}(\hat{s})  \sum_{q=u,d}   \frac{\left (m_{u} \frac{v_{\eta}}{v_{\rho}}- m_{d} \frac{v_{\rho}}{v_{\eta}} \right )}{m_{q}}   I_q  \ \sum_{q=u,d}  I_q^{*}           ,
\end{eqnarray}

\begin{eqnarray} 
\left (\frac{d \hat{\sigma}}{d\cos \theta} \right )_{pp \rightarrow Z H_{1}^{0}}^{Z(Z^{\prime})-H_{2}^{0}} & = & - \frac{g^{3}(g^{4}) \ \alpha_s^2   \ \Lambda_{Z(Z^{\prime}) \ ZH_{2}^{0}} \ v_{\eta} v_{\rho} \ \Pi \ \beta_{H_{i}^{0}}}{1024 \ \pi^{3} \ \hat{s}^{2} \  \cos^{2}_{\theta_{W}} v_{W}^{3}} {\it Re} \left[ \frac{\chi^{(2)}(\hat{s})}{(s- m_{Z'}^{2}+ im_{Z'} \Gamma_{Z'})}    \right.  \nonumber  \\
&& \left. \sum_{q=u,d}  m_{q} \left (m_{u} \frac{v_{\eta}}{v_{\rho}}- m_{d} \frac{v_{\rho}}{v_{\eta}} \right )  T_3^q \left( 1 + 2 \delta_q I_q  \right) \ \ \sum_{q=u,d}  I_q^{*} \right]          ,
\end{eqnarray}

\begin{eqnarray} 
\left (\frac{d \hat{\sigma}}{d\cos \theta} \right )_{pp \rightarrow Z H_{2}^{0}}^{H_{1}^{0}} & = &   \frac{g^{2} \ \alpha_s^2 \  v_{\eta}^{2} \ v_{\rho}^{2} \ \hat{s} \ \Omega \ \beta_{H_{i}^{0}}}{8192 \ \pi^{3}  v_{W}^{6} \  \cos^{2}_{\theta_{W}}} \ |\chi^{(1)}(\hat{s})|^{2}  \left|
\sum_{q=u,d}   2 \delta_q + \delta_q (4 \ \delta_q-1)  I_q  \right|^2   + \nonumber   \\
&& + \frac{g^{2} \ \alpha_s^2 \ v_{\eta}^{2} \ v_{\rho}^{2}  \ \Omega \ \beta_{H_{i}^{0}}}{4096 \ \pi^{3} \ \hat{s} \  \cos^{2}_{W}  v_{W}^{6}}   \ |\chi^{(1)}(\hat{s})|^{2} \left|\sum_{q=u,d} m_{q}^{2}  I_q  \right|^2     ,
\end{eqnarray}

\begin{eqnarray} 
\left (\frac{d \hat{\sigma}}{d\cos \theta} \right )_{pp \rightarrow Z H_{2}^{0}}^{H_{2}^{0}} & = &  \frac{g^{2} \ \alpha_s^2  \ (v_{\rho}^{2} - v_{\eta}^{2} )^{2} \ \hat{s} \ \Omega \ \beta_{H_{i}^{0}}}{131072 \pi^{3}  v_{W}^{6}  \  \cos^{2}_{\theta_{W}}}  \ |\chi^{(2)}(\hat{s})|^{2}  \left|
\sum_{q=u,d} \frac{\left(m_{u} \frac{v_{\eta}}{v_{\rho}}- m_{d} \frac{v_{\rho}}{v_{\eta}} \right)}{m_{q}} \left[ 2 \delta_q + \delta_q (4 \ \delta_q-1)  I_q  \right] \right|^2    \nonumber   \\
&& + \frac{g^{2} \ \alpha_s^2  \ (v_{\rho}^{2}-  v_{\eta}^{2} )^{2}  \ \Omega \  \beta_{H_{i}^{0}}}{65536 \ \pi^{3} \hat{s} \  v_{W}^{6} \ \cos^{2}_{W}} \ |\chi^{(2)}(\hat{s})|^{2} \left|\sum_{q=u,d}  m_{q} \left (m_{u} \frac{v_{\eta}}{v_{\rho}}- m_{d} \frac{v_{\rho}}{v_{\eta}} \right ) I_q  \right|^2           ,
\end{eqnarray}

\begin{eqnarray} 
\left (\frac{d \hat{\sigma}}{d\cos \theta} \right )_{pp \rightarrow Z H_{2}^{0}}^{Z^{\prime}-H_{1}^{0}} & = & - \frac{g^{4} \ \alpha_s^2   \ \Lambda_{ZZ'H_{2}^{0}} \ v_{\eta} v_{\rho} \ \Pi \ \beta_{H_{i}^{0}}}{1024 \ \pi^{3} \  \hat{s}^{2} \  \cos^{2}_{\theta_{W}} v_{W}^{3}} {\it Re} \left[ \frac{\chi^{(1)}(\hat{s})}{(s- m_{Z'}^{2}+ im_{Z'} \Gamma_{Z'})}    \right.  \nonumber  \\
&& \left. \sum_{q=u,d}  m_{q}^{2} \ T_3^q \left( 1 + 2 \delta_q I_q  \right) \ \ \sum_{q=u,d}  I_q^{*} \right]         ,
\end{eqnarray}

\begin{eqnarray} 
\left (\frac{d \hat{\sigma}}{d\cos \theta} \right )_{pp \rightarrow Z H_{2}^{0}}^{Z^{\prime}-H_{2}^{0}} & = & - \frac{g^{4} \ \alpha_s^2   \ \Lambda_{ZZ'H_{2}^{0}} \ (v_{\rho}^{2}- v_{\eta}^{2}) \ \Pi \ \beta_{H_{i}^{0}}}{4096 \ \pi^{3} \  \hat{s}^{2} \  \cos^{2}_{\theta_{W}} v_{W}^{3}} {\it Re} \left[ \frac{\chi^{(1)}(\hat{s})}{(s- m_{Z'}^{2}+ im_{Z'} \Gamma_{Z'})}    \right.  \nonumber  \\
&& \left. \sum_{q=u,d}   m_{q} \left (m_{u} \frac{v_{\eta}}{v_{\rho}}- m_{d} \frac{v_{\rho}}{v_{\eta}} \right ) \ T_3^q \left( 1 + 2 \delta_q I_q  \right) \ \ \sum_{q=u,d}  I_q^{*} \right]              ,
\end{eqnarray}
where in Eq. (25) are considered the contribution of $Z$ and $Z^{\prime}$ bosons, in Eq. (26) the contribution of $H_1^0$, in Eq. (27) the contribution of $H_2^0$, in Eq. (28) the contribution of interference of $Z Z^{\prime}$, in Eq. (29) the contribution of interference of $ZH_1^0$ and $Z^{\prime} H_1^0$, in Eq. (30) the contribution of interference of $H_1^0 H_2^0$ and in Eq. (31) the contribution of interference of $Z H_2^0$ and $Z^{\prime} H_2^0$, all these contributions are to produce the $Z H_1^0$. All other equations (30, 32, 33, 34, 35) are to produce the $Z H_2^0$, we point out that the term $\left (\frac{d \hat{\sigma}}{d\cos \theta} \right )_{pp \rightarrow Z H_{2}^{0}}^{H_{1}^{0}-H_{2}^{0}}$ is similar to $\left (\frac{d \hat{\sigma}}{d\cos \theta} \right )_{pp \rightarrow Z H_{1}^{0}}^{H_{1}^{0}-H_{2}^{0}}$, because the coupling constants $\Lambda_{H_{2}^{0}H_{2}^{0}Z} = - \Lambda_{H_{1}^{0}H_{1}^{0}Z}$ are equal. The sum runs over all generations, $T_3^q$ is the quark weak isospin  [$T_{3}^{u(d)} = +(-)1/2$], ${\it Re}$ stands for the real part of the expression.  The loop function $I_{i} \equiv I(\delta_{i} = m_{i}^{2} /\hat{s})$, is defined by 

\begin{eqnarray*}
I_i \equiv I_i (\delta_i) = \int_0^1 \frac{dx}{x} 
\ln \left[1 - \frac{(1-x)x}{\delta_i} \right] =   
\left \{ \begin{array}{l}
- 2 \left[ \sin^{-1}\left( \frac{1}{2 \sqrt{\delta_{i}}} \right)\right]^2
\; , \;\;\; \delta_i > \frac{1}{4} \\ \nonumber
\frac{1}{2} \ln^2 \left(\frac{r_+}{r_-}\right) - \frac{\pi^2}{2}  + i\pi 
\ln\left(\frac{r_+}{r_-}  \right)  \; , \;\;\; \delta_i < \frac{1}{4} ,
\end{array}
\right.
\label{ii}
\end{eqnarray*}
with, $r_\pm = 1 \pm (1 - 4 \delta_i)^{1/2}$ and $\delta_i = 
m_i^2/\hat{s}$. Here, $i = q$ stands for the particle (quark ) running in the loop.

We have also defined $\Delta$, $\Omega$ and $\Pi$ which are equal to:

\[
\Delta = 4  \hat{s} - \frac{\hat{u}^{2}}{2  m_{Z}^{2}}  + \frac{\hat{t} \ \hat{u}}{m_{Z}^{2}} - \frac{\hat{t}^{2}}{2  m_{Z^{2}}}
\]

\[
\Omega = \frac{\hat{s}^{2}}{4 m_{Z}^{2}} - \frac{\hat{s}}{2} + \frac{m_{Z}^{2}}{4}- \frac{\hat{s} \ m_{H_{1}^{0}}^{2}}{2 m_{Z}^{2}} - \frac{m_{H_{1}^{0}}^{2}}{2} + \frac{m_{H_{1}^{0}}^{4}}{4 m_{Z}^{2}}
\]

\begin{eqnarray*}
\Pi & = & -\frac{\hat{s}^{4}}{8 \ m_{Z}^{4}} + \frac{\hat{s}^{3}}{4 \ m_{Z}^{2}} -\frac{\hat{s}^{2} \hat{u}}{8 \ m_{Z}^{2}} -\frac{\hat{s}^{2} \hat{t}}{8 \ m_{Z}^{2}} + \frac{\hat{s}^{2}}{8} + \frac{3 \ \hat{s} \hat{u}}{8} + \frac{3 \ \hat{s} \hat{t}}{8} - \frac{\hat{s} \ m_{Z}^{2}}{4}  +\frac{\hat{s}^{3} \ m_{H_{i}^{0}}^{2}}{4 \ m_{Z}^{4}}       \nonumber  \\ 
&& +\frac{\hat{s}^{2} \ m_{H_{i}^{0}}^{2}}{4 \ m_{Z}^{2}} +\frac{\hat{s} \ \hat{u} \ m_{H_{i}^{0}}^{2}}{8 \ m_{Z}^{2}}  +\frac{\hat{s} \ \hat{t} m_{H_{i}^{0}}^{2}}{8 \ m_{Z}^{2}} - \frac{3 \ \hat{s} \ m_{H_{i}^{0}}^{2}}{4} -\frac{\hat{s}^{2} \ m_{H_{i}^{0}}^{4}}{4 \ m_{Z}^{4}}   \nonumber  \ , 
\end{eqnarray*}

The total cross section for the process $pp \rightarrow gg \rightarrow  Z  H_{i}$ is related to the subprocess $gg \rightarrow Z  H_{i}$ total cross section $\hat{\sigma}$ through  
\begin{equation}   
\sigma = \int_{\tau_{min}}^{1}  \int_{\ln{\sqrt{\tau_{min}}}}^{-\ln{\sqrt{\tau_{min}}}} d\tau dy  G\left(\sqrt{\tau}e^y, Q^2\right) G\left(\sqrt{\tau}e^{-y}, Q^2\right)   \hat{\sigma}\left(\tau, s\right), 
\end{equation}
where $G\left(x,Q^2\right)$ is the gluon structure function and $\tau_{min}$ is given above.

\section{RESULTS AND CONCLUSIONS}   

In the SM at the Tevatron, Higgs-boson production in association  with W or Z bosons, $pp \rightarrow W \ H_{1}^{0} /Z \ H_{1}^{0} + X$ is the most promissing discovery channel for a SM Higgs $H_{1}^{0}$ masses below about 130 GeV, where $H_{1}^{0} \rightarrow b \bar{b}$ decays are dominant \cite{carena}, while Higgsstrahlung is only marginal at the LHC.  \par 

We consider Higgs masses spanning in the range 115 GeV $\leq m_{H_{1}^{0}} \leq$ 800 GeV. Our talk will be subdivided into two distinct classes, if $m_{H_{1}^{0}}$ is less than 145 GeV or greater than 466 GeV, then ATLAS excludes Higgs boson masses above 145 GeV till 466 GeV, \cite{atlas1, atlas}. The best chance of discovering a SM Higgs at the LHC appears to be given by the following processes: \par 

$\bullet$ $gg \rightarrow H_{1}^{0} \rightarrow \gamma \gamma$, where the highest rates for this channel occur in the region 80-150 GeV and the combination of a rising branching ratio and a falling cross section \cite{moretti}, yields rates which are constant over the above range. \par  

$\bullet$ $q \bar{q}{^\prime} \rightarrow WH_{1}^{0} \rightarrow \ell \nu_{\ell} \gamma \gamma$, these channels are different with respect to the case of the direct $gg \rightarrow H_{1}^{0} \rightarrow \gamma \gamma$ production and decay. On the one hand owing to a larger number of reducible ($W \gamma j$ with the pions in the jet giving hard photons and $Z \gamma$ with a $\ell^{\pm}$ from the Z decay faking a photon) and irreducible  ($W \gamma \gamma$) backgrounds. On the other hand owing to the isolated hard lepton from the W decays that allows for a strong reduction of the backgrounds \cite{moretti1}. Must be here noticed that unlike the signals from leptonic decay modes, those from hadronic decay channels undergo mostly QCD background processes. \par 

$\bullet$ $H \rightarrow b \bar{b}$; $q \bar{q}{^\prime} \rightarrow WH \rightarrow \ell \bar{\nu}_{\ell} b \bar{b}$;  $gg, q \bar{q}{^\prime} \rightarrow  t \bar{t} H \rightarrow q \bar{q} q \bar{q} WW \rightarrow q \bar{q} q \bar{q} \ell \bar{\nu}_{\ell} X$, where the first decay mode is the dominant in the range 80 GeV $\leq m_{H_{1}^{0}} \leq $ 130 GeV. All these decay modes use the techniques of flavor identification of b-jets, thereby reducing the huge QCD background from light quarks and gluon jets. The chances to tag the Higgs boson in association with a W or a $t \bar{t}$ pair are $W \rightarrow \ell \bar{\nu}_{\ell}$ and $t \bar{t} \rightarrow \ell \bar{\nu}_{\ell} X$. The lepton is usually at high $p_{T}$ and isolated. The expected signatures would then be $\ell b \bar{b} X$ from $WH_{1}^{0}$ and $b \bar{b} b \bar{b} \ell X$ from $t \bar{t}H_{1}^{0}$. Higgs signal in the $b \bar{b}$ channel would appear as a peak in the invariant mass distribution of b-quark pairs. \par   

The backgrounds to the channel $WH_{1}^{0} \rightarrow \ell b \bar{b} X$ that we have considered are: $W b \bar{b}$, $Z b \bar{b}$, $WZ$, $ZZ$, $q \bar{q}{^\prime} \rightarrow t \bar{b}, t \bar{t}$, $q g \rightarrow t \bar{b} q^{\prime}$, $Wjb$ and $Wjj$. The dominant irreducible backgrounds are $Wb \bar{b}$, $WZ$ with $Z \rightarrow b \bar{b}$. The dominant reducible background comes from $Wjj$ production, through $Wgg$, $Wgq$ and $Wq \bar{q}$ in which the two jets are faked as b-quarks and $Wjb$ for one jet only. The irreducible background in $t \bar{t} Z$ events is small with respect to the signal $t \bar{t} H_{1}^{0}$. \par      

$\bullet H \rightarrow Z \ Z \rightarrow 4 \ \ell$ this channel with $\ell = e$ or $\mu$ nicknamed as the `gold-plated' has recibed much attention in the literature because of the opportunity for a fully reconstructed Higgs signal. This channel provides the best chances for Higgs detection over a large interval of the Higgs mass range, between $\simeq$ 120 and 800 GeV. Its signature is relatively clean, especially if compared to the difficulties encountered for the $H_{1}^{0} \rightarrow \gamma \gamma$ and $H_{1}^{0} \rightarrow b \bar{b}$ cases, however, this process ocurrs with a small branching ratio, $B= \frac{1}{3} \times (0.034)^{2}=3.9 \times 10^{-4}$, that the predicted events rates are small. In the mass region $m_{H_{1}^{0}} \leq 2 m_{Z}$ the main backgrounds to the four lepton signal come from $t \bar{t} \rightarrow b \bar{b} W^{+}W^{-}$, $Zb\bar{b}$ and $ZZ$ production, where the last is irreducible. In the first two backgrounds two of the leptons come from semi-leptonic b-decays, while the remaining two from the decays of the massive vector bosons, that is $W^{+}W^{-} \rightarrow \ell^{+} \ell^{-} X$ and $Z \rightarrow \ell^{+} \ell^{-}$. In the case of the high mass region, when the Higgs decay  $H_{1}^{0} \rightarrow ZZ$ can occur, the only significant background, after appling the invariant mass cut $m_{\ell^{+} \ell^{-}} \simeq m_{Z}$ on both the lepton-antilepton pair, is the continuum $ZZ$ production \cite{moretti}. The Higgs signal would appear as a resonance, which has a maximum at $m_{4\ell} = 2m_{Z}$. \par 

$\bullet$  $H_{1}^{0} \rightarrow ZZ \rightarrow \ell \bar{\ell} \nu \bar{\nu}$, this channel should also be considered as it is enhanced relative to the gold-plated decay mode by a factor of six \cite{cahn}, owing to its BR which is six times larger. This channel can give good chances for Higgs detection for high values of $m_{H_{1}^{0}} \geq 700$ GeV. While this process may at first seem an unlikely channel since the second Z is not detected, the signature reveals a Z with two high $p_{T}$ leptons and high missing transverse energy from the other Z. The main backgrounds to this channel are the continuum production of ZZ and WZ. Other possible reducible backgrounds are $q \bar{q} \rightarrow ZW \rightarrow \ell^{+} \ell^{-} \ell^{\prime} \nu$ where the charged leptons $\ell^{\prime} = e, \mu, \tau $ escapes detection. Another reducible backgrounds arises from Z+ jets, where enough of the jets escape detection to leave large missing $p_{T}$ and the production of $t \bar{t}$, but to suppress this background the additional cut $|m_{Z}- m_{\ell^{+} \ell^{-}} \leq |$ 6 GeV must be applied. \par            

In this work we have calculated contributions regarding to the Drell-Yan and gluon-gluon fusion in 3-3-1 model. We present the cross section for the process $pp \rightarrow Z H_{i}$ involving the Drell-Yan mechanism and the gluon-gluon fusion, to produce such Higgs bosons for the LHC. In all calculations was taken for the parameters, the VEV and masses the following representative values, \cite{TO96, cnt2} : $\lambda_{1} =-0.20$,  $\lambda_{2}=1.02$, $\lambda_{3}=-\lambda_{6}=-1$, $\lambda_{4}= 2.98$ $\lambda_{5}=-1.57$, $\lambda_{7} =-2$, $\lambda_{8}=-0.42$,  $v_{\eta}=195$ GeV, $\lambda_{9}=-0.9(-0.76)$, for $v_{\chi}=1000(1500)$ GeV, these parameters are used to estimate the values for the particles masses which are given in Table I, it is to notice that the value of $\lambda_{9}$ was chosen this way in order to guarantee the approximation $-f \simeq v_{\chi}$ \cite{TO96, cnt2}.

\begin{table}[h]
\caption{\label{tab1} Values for the particle masses used in this work. All the values in this Table are given in GeV.}
\begin{ruledtabular}
\begin{tabular}{c|ccccccccccccccc}
f & $v_{\chi}$ & $m_E$ & $m_M$ & $m_T$ & $m_{H^{\pm \pm}}$ & $m_{H_3^0}$ & $m_{h^0}$ & $m_{H_1^0}$ & $m_{H_2^0}$ & $m_{H^\pm_2}$ & $m_V$ & $m_U$   \\
\hline
-1008.3 & 1000 & 148.9 & 875 & 2000 & 500  &  2000 & 1454.6  & 126  & 1017.2 & 183 & 467.5 & 464   \\
-1499.7 & 1500   & 223.3  & 1312.5  & 3000  & 500  & 3000 & 2164.3  & 126 & 1525.8 & 285.2 &  694.1 & 691.7   \\
\end{tabular}
\end{ruledtabular}
\end{table}


\begin{ruledtabular}
\hskip 11.5cm \begin{tabular}{cccc}
\hskip 11.5cm $m_{Z^\prime}$ & $m_{J_1}$ & $m_{J_2}$ & $m_{J_3}$ \\
\hline 
\hskip 11.5cm  1707.6  & 1000  & 1410 & 1410    \\
\hskip 11.5cm 2561.3 & 1500 & 2115 & 2115  \\
\hskip 11.5cm \end{tabular}
\end{ruledtabular}



\subsection{The Higgs $H_{1}^{0}$}

We may have very strong hints of the standard lightweight Higgs ($m_{H_{1}^{0}} \propto$ 125 GeV), with the b-tagging results from CDF and DZero of the TEVATRON and with the two-photon and four-lepton results from ATLAS and CMS of the LHC, at the level of 5 sigma, in the mass region around 125-126 GeV \cite{cms, atlas11}. We interpret this to be due to the production of a previously unobserved particle with a mass of around 125 GeV. At the end of 2012, CMS and ATLAS updated all the Higgs decay channels  for 13 fb$^{-1}$ and 8 TeV, except that CMS choose not to publish the diphoton result because it was smaller than expected. The hope is that they will get full results at 20 fb$^{-1}$ and 8 TeV and they will enable to elucidate further the nature of this newly observed particle. In addition to the cross-sections we can hope for an update to the tests of spin parity on the Higgs boson. This will be the final step to declare that the Higgs-Very-Like-Boson is indeed the standard Higgs-Boson. Even so, until we can not measure a signal of a Higgs in a completely convincing way, we have a right to make the phenomenology analysis on the intermediate mass Higgs. \par 

So in Fig. 3 and 4, we show the cross section $pp \rightarrow Z H_{1}^{0}$ at $\sqrt{s}=$ 8 (14) TeV, these processes will be studied in two cases, the one is for the vacuum expectation value $v_{\chi}=1000$ GeV and the other is for $v_{\chi}=1500$. Considering that the expected integrated luminosity for the LHC collider that will be reach is of order of $20 \ fb^{-1} (300 $ fb$^{-1}$) then the statistics for $v_{\chi}=1000$   gives a total of   $\simeq 2.6 \times 10^4( 2.0 \times 10^4) (2.1 \times 10^6(1.7 \times 10^6 ))$ events per year for Drell-Yan and $\simeq 360(328) (2.9 \times 10^5(2.6 \times 10^5 ) ) $ events per year for gluon-gluon fusion, if we take the mass of the Higgs boson $m_{H_{1}^{0}}= 650(800)$ GeV respectively ($\Gamma_{H_{1}^{0}} = 214, 551.7$ GeV). Here we must clarify that the first two number of events  $\simeq 2.6 \times 10^4 ( 2.0 \times 10^4)$ are relative to Drell-Yan and the other two $\simeq 360(328)$ to gluon-gluon fusion and correspond to 8 TeV and the other pair $\simeq 2.1 \times 10^6(1.7 \times 10^6 )$ are relative to Drell-Yan and $\simeq 2.9 \times 10^5(2.6 \times 10^5 )$ to gluon-gluon fusion and correspond to 14 TeV for the LHC, respectively.  \par 

Next, we multiply the production cross sections by the respective branching ratios to obtain event rates for various channels. Considering that the signal for  $ZH_{1}^{0}$ production for  $m_{H_{1}^{0}}= 650(800)$ GeV will be $ Z \rightarrow \ell^{+} \ell^{-}$ and $H_{1}^{0} \rightarrow ZZ \rightarrow \ell^{+} \ell^{-} \ell^{+} \ell^{-}$, then taking into account that the branching ratios for both particles would be $B(Z \to \ell^{+} \ell^{-}) = 3.4 \ \% $ and  $B(H^{0}_{1} \to ZZ) = 29.0 (30.9) \ \% $, see Fig. 5, we would have approximately $ \simeq 0(0) (24(21))$ events per year.  In respect to gluon-gluon fusion we will have  $ \simeq 0(0)(3(4))$ events per year to produce the same particles. We will remember that the  first two number of events  $\simeq 0(0)$ refer to Drell-Yan and the other two $\simeq 0(0)$ to gluon-gluon fusion and correspond to 8 TeV and the other pair $\simeq 24 (21)$ refer to Drell-Yan and $\simeq 3(4)$ to gluon-gluon fusion and correspond to 14 TeV for the LHC, respectively.  \par 

Regarding the vacuum  expectation  value $v_{\chi}=1500$ GeV for the same masses of $m_{H_{1}^{0}}= 650(800)$ ($\Gamma_{H_1^0} = 212.2, 551.4$ GeV) it  will give a total of  $\simeq 4.0 \times 10^3( 3.6 \times 10^3) (7.1 \times 10^5(6.6 \times 10^5 ))$  events per year to produce $H_{1}^{0}$ for Drell-Yan. In respect to gluon-gluon fusion we will have  $ \simeq 2(2)(9.0 \times 10^3(8.9 \times 10^3 ))$ events per year to produce the same particles. Taking into account the same signal as above, that is $B(Z \to \ell^{+} \ell^{-}) = 3.4 \ \% $ and  $B(H^{0}_{1} \to ZZ) = 29.2 (30.9) \ \% $, see Fig. 6, we would have approximately $ \simeq 0(0) (8(8))$ events per year for Drell-Yan and $\simeq 0 (0)(0 (0))$ for gluon-gluon fusion. Comparing these signatures with the standard model background, like $pp \rightarrow ZZZ$ and using COMPHEP \cite{pukhov}, we have that a cross section at LO is $3.8 \times 10^{-3} ( 9.6 \times 10^{-3})$ pb, where the fist value is for $\sqrt{s}=8$ and the second for  $\sqrt{s}=14$ TeV, respectively. Considering now the same signature as above we have $\simeq 0(0)$ events for the background. So we will have a total of $\simeq 27$ events per year for the signal for $m_{H_{1}^{0}} = 650$ GeV and $\simeq 25$ events per year for  $m_{H_{1}^{0}} = 800$ GeV, for $v_\chi = 1000$ GeV and for $\sqrt{s}=14$ TeV and $\simeq 8(8)$ events per year for  $m_{H_{1}^{0}} = 650(800)$ for  $v_\chi = 1500$ GeV and for $\sqrt{s}=14$ TeV. To extract the leptonic signal from the background we impose the Z window cut where the invariant mass of opposite-charge leptons must be far from the Z mass: $|m_{\ell^{+} \ell^{-}} - m_{Z}| >$ 10 GeV, this removes events where the leptons come from Z decay \cite{cheng}. Then the Higgs signal would appear as a broad Breit-Wigner resonance, which has a maximum at $m_{4 \ell} = 2 m_{Z}$. \par 
 
Taking another signal such as $H_{1}^{0} Z \rightarrow W^{+}W^{-} Z$, and taking into account that the branching ratios for these  particles would be $B(H^{0}_{1} \to W^{+}W^{-}) = 59.1 (62.3)  \ \% $,  see Fig. 5 and $B(Z \to \ell^{+} \ell^{-}) = 3.4 \ \% $, for the mass of the Higgs boson $m_{H_{1}^{0}}= 650(800)$ GeV, $v_{\chi}=1000$ GeV, and that the particles $W^{+} W^{-}$ decay into $e^{+} \nu$ and $e^{-} \bar{\nu}$, whose branching ratios for these particles would be  $B(W^{\pm} \to e^{\pm} \nu = 10,75 \ \%) $, then  we would have approximately $\simeq 6 (4)(487 (414))$ events per year for Drell-Yan. In respect to gluon-gluon fusion we will have  $ \simeq 0(0)(67 (65))$ events per year to produce the same particles. Regarding the vacuum  expectation value $v_{\chi}=1500$ GeV and considering that the  branching ratios for $H_{1}^{0}$ would be $B(H^{0}_{1} \to W^{+}W^{-}) = 59.4 (62.4)  \ \% $,  see Fig. 6 and taking the same parameters and branching ratios for the same particles given above, then we would have for $m_{H_{1}^{0}}= 650(800)$ a total of  $ \simeq 0(0)(166 (161))$ events of $H_{1}^{0}$ produced per year for Drell-Yan and in respect to gluon-gluon fusion the number of events per year will be $ \simeq 0(0)(2 (2))$. \par 

Taking the largest standard model background, like $pp \rightarrow W^{+} W^{-} Z$ and using CompHep \cite{pukhov} we have that this yields a cross section of $3.8 \times 10^{-2}(9.8 \times 10^{-2})$ pb at LO for $\sqrt{s}=8(14)$ TeV respectively. Taking into account the $\ell^{+} \ell^{-} e^{+} e^{-} X$ backgrounds we have a total of $\simeq 0(12)$ events, so we will have a total of 
$\simeq 554$ events per year for the signal for $m_{H_{1}^{0}} = 650$ GeV and $\simeq 479$ events per year for  $m_{H_{1}^{0}} = 800$ GeV, both for $v_\chi = 1000$ GeV and for $\sqrt{s}=14$ TeV and $\simeq 168(163)$ events per year for  $m_{H_{1}^{0}} = 650(800)$ for  $v_\chi = 1500$ GeV and for $\sqrt{s}=14$ TeV and  $\simeq 12$ events for the background. To make clear these signals we  must isolate the hard lepton from the W with $p_{T}^{\ell} >$ 20 GeV and adopt the cut on the missing transverse momentum ${p\!\!\slash}_{T} >$ 20 GeV. \par  

The $H_{1}^{0} Z$ will also decay into  $H^{\pm \pm} H^{\mp \mp} Z$, and taking into account that the branching ratios for these  particles would be $B(H^{0}_{1} \to  H^{\pm \pm} H^{\mp \mp}) = 6.6 (2.8)  \ \% $, see Fig. 5 and $B(Z \to b \bar{b}) = 15.2 \ \% $ for the mass of the Higgs boson $m_{H_{1}^{0}}= 650(800)$ GeV and $v_{\chi}=1000$ GeV, and that the particles $H^{\pm \pm}$ decay into  $e^{\pm} E^{\pm}$, whose branching ratios for these particles would be  $BR(H^{\pm \pm} \to e^{\pm} E^{\pm}) = 44.0(31.1) \ \%$ \cite{quasi}, these branching ratios are in good agreement with the upper cinematic limit of Higgs mass \cite{cnt2}, which in this case are $m_{H^{\pm \pm}}=325(400)$ GeV, then we would have approximately $\simeq 50 (8)(4.1 \times 10^{3} (700 ))$ events per year for Drell-Yan. Must be taken into account that E is the new heavy lepton and we assume that ($P=E, M, T$) \cite{TO96,pphmm}. In respect to gluon-gluon fusion we will have  $ \simeq 0(0)(5.6 \times 10^{2} (107))$ events per year to produce the same particles.  \par 

With respect to $v_{\chi}=1500$ and doing all the analysis given above for the mass of Higgs boson $m_{H_{1}^{0}}= 650(800)$, considering that $B(H^{0}_{1} \to  H^{\pm \pm} H^{\mp \mp}) = 6.1 (2.6)  \ \% $,  see Fig. 6  and $B(Z \to  b \bar{b}) = 15.2 \ \% $ and $BR(H^{\pm \pm} \to e^{\pm} E^{\pm}) = 50.0(31.1) \ \%$  \cite{quasi}, then it will take a total of $\simeq 10(2)(1.6 \times 10^{3} (252))$ events per year for Drell-Yan and $\simeq 0(0)(21(3 ))$ events for gluon-gluon fusion. In this way we have as signal four leptons and Z would appear in the invariant mass distribution of b-quark pairs, therefore, if we observe this signal, additionally to $H_{1}^{0}$ we observe also the doubly charged Higgs boson and heavy leptons, consequently it is a very striking and important signal. So we will have for $\sqrt{s}=8$ TeV a total of $\simeq 50(8)$ events for signal $b \bar{b} e^{+} E^{-} e^{-} E^{+}$, for $m_{H_1^0}=650(800)$ and for $v_\chi = 1000$ GeV and $\simeq 10(2)$ events for $v_\chi = 1500$ GeV and for the same parameters cited above. It is important to be noticed if the standard lightweight Higgs ($m_{H_{1}^{0}} \propto$ 125 GeV) has not been discovered, this channel of decay cited above is the most important. For $\sqrt{s}=14$ TeV we will have a total of $\simeq 4.6 \times 10^{3}(807)$ events per year for $m_{H_1^0}=650(800)$, for $v_\chi = 1000$ GeV and $\simeq 1.6 \times 10^{3}(255)$ events per year for $m_{H_1^0}=650(800)$ and for  $v_\chi = 1500$ GeV to produce the same signal. \par 

The main background to this signature is the ZZZ production which gives a cross section at LO, using COMPHEP \cite{pukhov}, of  $3.8 \times 10^{-3} ( 9.6 \times 10^{-3})$ pb, for $\sqrt{s}=8(14)$ TeV respectively. We consider the irreducible background, that is $ZZZ \rightarrow b \bar{b} \ell^{+} \ell^{-} \ell^{+} \ell^{-}$, then we have $\simeq 0(0)$ events, therefore to observe these signals with the Z which decay into $b \bar{b}$ channel, first we must select the $b \bar{b}$ channel out of the huge QCD backgrounds of quark and gluon jets by using the b-tagging capabilities of vertex detectors, later we apply the Z window cut $|m_{\ell^{+} \ell^{-}} - m_{Z}| >$ 10 GeV, which removes events where the leptons come from Z decay \cite{cheng}. Therefore the Higgs signal would appear as a resonance, which has a maximum at  $m_{4 \ell} = 2 m_{H^{\pm \mp}}$. \par 

Taking another channel of decay such as  $H_{1}^{0} Z \rightarrow t \bar{t} \ \ell^{+} \ell^{-}$, and considering that the branching ratios for these particles would be  $B(H^{0}_{1} \to  t \bar{t}) = 5.1 (3.9)  \ \% $,  see Fig. 5 and $B(Z \to \ell^{+} \ell^{-}) = 3.4 \ \% $ for the mass of the Higgs boson $m_{H_{1}^{0}}= 650(800)$ GeV, $v_{\chi}=1000$ GeV, and that the  $t \bar{t}$ particles decay into $ b \bar{b} W^{+} W^{-}$, whose branching ratios for these particles would be $B(t \to b W) = 99.8 \ \% $ followed by leptonic decay of the boson W, that is  $B(W \to e \nu) = 10.75 \ \% $, then our signal will be $b \bar{b} e^{+} e^{-} \ell^{+} \ell^{-} X$ and we would have approximately $\simeq 0 (0)(42 (26))$ events per year for Drell-Yan for the parameters listed above. \par 

Regarding to gluon-gluon fusion we will have  $ \simeq 0(0)(6 (4))$ events per year to produce the same particles. Considering the vacuum  expectation value $v_{\chi}=1500$ GeV and the branching ratios $B(H_{1}^{0} \rightarrow t \bar{t}) = 5.2 (3.9) \ \% $, see Fig. 6 and taking the same parameters and branching ratios for the same particles given above, then we would have for $m_{H_{1}^{0}}= 650(800)$ a total of  $ \simeq 0(0)(14 (10))$ events of $H_{1}^{0}$ produced per year and in respect to gluon-gluon fusion the number of events per year will be $ \simeq 0(0)(0 (0))$. In this way we have a total of $\simeq 0 (0)(48 (30))$ events per year for $\sqrt{s}=8(14)$ TeV, for  $m_{H_{1}^{0}}= 650(800)$ GeV, for  $v_{\chi}=1000$ GeV and $\simeq 0 (0)(14 (10))$ events per year for $\sqrt{s}=8(14)$ TeV, for  $m_{H_{1}^{0}}= 650(800)$ GeV and for $v_{\chi}=1500$ GeV. Taking the irreducible background $t \bar{t} Z  \rightarrow b \bar{b} e^{+} e^{-} \ell^{+} \ell^{-} X$, and using CompHep we have that a cross section at LO is $\simeq 2.3 \times 10^{-1} (1)$ pb for $\sqrt{s}=8(14)$ TeV. Then the number of events will be  $ \simeq 1(117)$, in this case we have that the number of backgrounds is greater than the number of signals and consequently the statistical significance is $\simeq 4.4(2.8) \sigma$ for $m_{H_{1}^{0}}= 6500(800)$ GeV, for $v_{\chi}=1000$ GeV, for $\sqrt{s}=14$ TeV, and  $\simeq 1.3(0.9) \sigma$ for  $m_{H_{1}^{0}}= 650(800)$ GeV, for $v_{\chi}=1500$ GeV and for $\sqrt{s}=14$.  Must be imposed a set of kinematic cuts on all the missing transverse momentum and the lepton momenta to improve the statistical significance of a signal, must be isolated  a hard lepton from the $W$ decay with $p_{T}^{\ell}>$ 20 GeV, put the cut on the missing transverse momentum ${p\!\!\slash}_{T} >$ 20 GeV and apply the Z window cut $|m_{\ell^{+} \ell^{-}} - m_{Z}| >$ 10 GeV, which removes events where the leptons come from Z decay \cite{cheng}.


\subsection{The Higgs $H_{2}^{0}$}

The Higgs $H_{2}^{0}$ in 3-3-1 model is not coupled to a pair of standard bosons, it couples to quarks, leptons, Z Z', Z'Z' gauge bosons,  $H_{1}^{-} H_{1}^{+}$, $H_{2}^{-} H_{2}^{+}$, $h^{0}   h^{0}$, $H_{1}^{0}  H_{3}^{0}$ higgs bosons, $V^{-}V^{+}$ charged bosons, $U^{--} U^{++}$ double charged bosons, $H_{1}^{0} Z$, $H_{1}^{0} Z'$ bosons and $H^{--} H^{++}$ double charged Higgs bosons \cite{cieton2}. The Higgs $H_{2}^{0}$ can be much heavier than 1018 GeV for $v_\chi = 1000$ GeV, and 1526 for $v_\chi = 1500$GeV, so the Higgs $H_2^{0}$ is a hevy particle. The coupling of the $H_{2}^{0}$ with $H_{1}^{0}$ contributes to the enhancement of the total cross section $\it{via}$ the Drell-Yan and gluon-gluon fusion. \par  

In Fig. 7 and 8, we show the cross section $pp \rightarrow Z H_{2}^{0}$, these processes will also be studied in two cases, for the vacuum expectation value $v_{\chi}=1000$ GeV and for $v_{\chi}=1500$. Considering the expected integrated luminosity for the LHC collider given above, then the analysis for $v_{\chi}=1000$   gives a total of $\simeq 1.7 \times 10^3( 7.5 \times 10^2)(2.7 \times 10^{5} (1.4 \times 10^{5}))$ events per year for Drell-Yan and $\simeq 57(45)(9.5 \times 10^{4} (7.7 \times 10^{4}))$ events per year for gluon-gluon fusion, if we take the mass of the Higgs boson $m_{H_{2}^{0}}= 1100(1300)$ GeV ($\Gamma_{H_2^0}= 442.9, 715$ GeV). These values are in accord with the Table I. It must be noticed that must take care with large Higgs masses, as the width approaches the value of the mass itself for a very heavy Higgs and one looses the concept of resonance. Remember that the first two number of events $\simeq 1.7 \times 10^3( 7.5 \times 10^2)$ are relative to Drell-Yan and the other two $\simeq 57(45)$ to gluon-gluon fusion and correspond to 8 TeV and the other pair $\simeq (2.7 \times 10^{5} (1.4 \times 10^{5}))$ are relative to Drell-Yan and $\simeq (9.5 \times 10^{4} (7.7 \times 10^{4}))$ to gluon-gluon fusion and correspond to 14 TeV for the LHC, respectively. \par 

To obtain event rates we multiply the production cross sections by the respective branching ratios. Considering that the signal for  $H_{2}^{0}Z$ production for  $m_{H_{2}^{0}}= 1100(1300)$ GeV and $v_{\chi}=1000$ GeV will be $H_{2}^{0} Z \rightarrow  Z H_{1}^{0} Z$, and taking into account that the branching ratios for these particles would be $B(H^{0}_{2} \to Z H_{1}^{0}) = 46.2 (47.3)  \ \% $ and $B(Z \to  b \bar{b}) = 15.2 \ \% $, see Fig. 9, and that the particles $H_{1}^{0}$ decay into  $W^{+} W^{-}$, and taking into account that the branching ratios for these  particles would be $B(H^{0}_{1} \to W^{+} W^{-}) = 64.6 (63.2)  \ \% $, see Fig. 9, and then the $W^{+}$ decay into $\ell^{+} \nu$ and $W^{-}$ into $\ell^{-} \bar{\nu}$ whose branching ratios for these particles would be  $BR(W  \to \ell \nu) = 10.8 \ \%$, consequently we would have approximately  $ \simeq 0(0)( 22 (12))$  events per year for Drell-Yan and $\simeq 0(0)(8 (6 ))$ for gluon-gluon fusion for the signal $b\bar{b} b \bar{b} \ell^{+} \ell^{-} X$. \par 

With respect to vacuum  expectation  value $v_{\chi}=1500$ GeV for the masses of $m_{H_{2}^{0}}= 1600(1800)$ ($\Gamma_{H_2^0}= 1303, 1838$ GeV) it  will give a total of  $\simeq 228(121 ) (6.8 \times 10^4(4.4 \times 10^4 ))$  events per year to produce $H_{2}^{0}$ for Drell-Yan and in respect to gluon-gluon fusion we will have  $ \simeq 0(0)(2.9 \times 10^3(2.5 \times 10^3 ))$ events per year to produce the same particles. Taking into account the same signal as above, that is $b\bar{b} b \bar{b} \ell^{+} \ell^{-} X$ and considering that the branching ratios would be $B(H^{0}_{2} \to Z H_{1}^{0}) = 48.3 (48.7)  \ \%$, $B(H^{0}_{1} \to W^{+} W^{-}) = 65.7 (65.1)  \ \% $, $B(Z \to  b \bar{b}) = 15.2 \ \% $ and $BR(W  \to \ell \nu) = 10.8 \ \%$ we would have approximately $ \simeq 0(0) (6(4))$ events per year for Drell-Yan and $\simeq 0 (0)(0 (0))$ for gluon-gluon fusion. The main background to this signal is $t \bar{t} Z \rightarrow b\bar{b} b \bar{b} \ell^{+} \ell^{-} X$, which cross section at LO is $\simeq 1$ pb for $\sqrt{s}=14$ TeV. Then we have a total of $ \simeq 530$ events for the background and $\simeq 30(18)$ events for the signal for $m_{H_{2}^{0}}= 1100(1300)$ GeV, for $\sqrt{s}=14$ TeV and for $v_{\chi}=1000$, by other side for $v_{\chi}=1500$ the number of events for the signal is insignificant. \par 

Therefore we have that the statistical significance is $\simeq 1.3(0.8) \sigma$ for $m_{H_{2}^{0}}= 1100(1300)$ GeV, that is a low probability to detect
the signals. The improvement will be significant if we consider a luminosity $\simeq 10$ times higher than original LHC design, that is what we are awaiting to happen for 2025, then we will have $\simeq 300(180)$ events for the signals for $m_{H_{2}^{0}}=1100(1300)$ GeV, for $\sqrt{s}=14$ TeV and for $v_{\chi}=1000$, which corresponds to have a 5$\sigma$ discovery in the $b\bar{b} b \bar{b} \ell^{+} \ell^{-} X$ final state, by other side for $v_{\chi}=1500$ we have $60(40)$ events for $m_{H_{2}^{0}}= 1600(1800)$ GeV, for $\sqrt{s}=14$ TeV and which corresponds to 2.6(1.7)$\sigma$, for this last scenario the signal is too small to be observed even with 3000 $fb^{-1}$. To extract the signal from the background we must select the $b \bar{b}$ channel using the techniques of b-flavour identification, thus reducing the huge QCD backgrounds of quark and gluon jets, later the Z which coming together with the $H_{2}^{0}$ and the other Z coming from the decay of $H_{2}^{0}$ would appear as a peak in the invariant mass distribution of b-quark pairs. The charged lepton track from the $W$ decay and the cut on the missing transverse momentum ${p\!\!\slash}_{T} >$ 20 GeV allows for a very strong reduction of the backgrounds. \par 

The $H_{2}^{0} Z$ will also decay into  $t \bar{t} \ \ell^{+} \ell^{-}$, and consider that the branching ratios for these particles would be  $B(H^{0}_{2} \to  t \bar{t}) = 6.0 (4.5)  \ \% $,  see Fig. 9 and $B(Z \to \ell^{+} \ell^{-}) = 3.4 \ \% $ for the mass of the Higgs boson $m_{H_{2}^{0}}= 1100(1300)$ GeV, $v_{\chi}=1000$ GeV, and that the particles $t \bar{t}$ decay into $ b \bar{b} W^{+} W^{-}$, whose branching ratios for these particles would be  $B(t \to b W) = 99.8 \ \% $, followed by leptonic decay of the boson W, that is  $B(W \to e \nu) = 10.75 \ \% $, then  we would have approximately $\simeq 0 (0)(6 (2))$ events per year. Regarding to gluon-gluon fusion we will have  $ \simeq 0(0)(2 (1))$ events per year to produce the same particles. Considering the vacuum  expectation value $v_{\chi}=1500$ GeV and the branching ratios $B(H_{2}^{0} \rightarrow t \bar{t}) = 3.1 (2.5) \ \% $, see Fig. 10 and taking the same parameters and branching ratios for the same particles given above, then we would have for  $m_{H_{2}^{0}}= 1600(1800)$ a total of  $ \simeq 0(0)(2 (1))$ events of $H_{2}^{0}$ produced per year and in respect to gluon-gluon fusion the number of events per year will be $ \simeq 0(0)(0 (0))$. Taking again the irreducible background $t \bar{t} Z  \rightarrow b \bar{b} e^{+} e^{-} \ell^{+} \ell^{-} X$, and using CompHep we have that a cross section at LO is $\simeq 1$ pb for $\sqrt{s}=14$ TeV, which gives $ \simeq 117$ events. We consider only the events for the signal for  $v_{\chi}=1000$, which gives $\simeq 8(2)$, for $v_{\chi}=1500$, the number of events is insignificant. \par 

Then we have that the statistical significance is $\simeq 0.7(0.3) \sigma$ for $m_{H_{2}^{0}}= 1100(1300)$ GeV, for $v_{\chi}=1000$ GeV, for $\sqrt{s}=14$ TeV. For this scenario the signal significance is smaller than 1$\sigma$ and discovery can not be accomplished unless the luminosity will be improved. So, if we enhance the integrated luminosity up to 3000 $fb^{-1}$, then we will have $\simeq 60(30)$ events for the signals for $m_{H_{2}^{0}}=1100(1300)$ GeV, for $\sqrt{s}=14$ TeV and for $v_{\chi}=1000$, which corresponds to have a 5.6(2.8)$\sigma$ discovery in the $b \bar{b} e^{+} e^{-} \ell^{+} \ell^{-} X$ final state and for $v_{\chi}=1500$ the signal will be not visible in this channel. We impose the following cuts to improve the statistical significance of a signal, i. e. we isolate a hard lepton from the $W$ decay with $p_{T}^{\ell}>$ 20 GeV, put the cut on the missing transverse momentum ${p\!\!\slash}_{T} >$ 20 GeV and apply the Z window cut $|m_{\ell^{+} \ell^{-}} - m_{Z}| >$ 10 GeV, which removes events where the leptons come from Z decay \cite{cheng}. However, all this scenarios can only be cleared by a careful Monte Carlo work to determine the size of the signal and background. \par   

In summary, we showed in this work that in the context of the 3-3-1 model the signatures for neutral Higgs bosons can be significant in LHC collider. Our study indicates the possibility of obtaining a clear signal of these new particles through its different modes of decay. If this model is realizable in the nature, certainly new particles will appear such as $H_{1}^{0}, H_{2}^{0}, Z^{\prime}, P^{\pm}, H^{\pm \pm}$ in the context of this study. \par

\acknowledgments
{MDT is grateful to the Instituto de F\'\i sica Te\'orica for hospitality UNESP, the Brazilian agencies CNPq for a research grant, and FAPESP for financial support (Processo No. 2009/02272-2).}

\newpage
\newpage

\begin{center}
FIGURE CAPTIONS
\end{center}

  
{\bf Figure 1}: Feynman diagrams for production of neutral Higgs {\it via} Drell-Yan process.
 
{\bf Figure 2}: Feynman diagrams for production of neutral Higgs {\it via} gluon-gluon fusion.  

{\bf figure 3}:Total cross section for the process $p \ p \to
Z H_{1}^{0}$ as a function of $m_{H_{1}^{0}}$ for a
$v_{\chi}=1.0(1.5)$ TeV at $\sqrt{s} = 8$ TeV for Drell-Yan-
solid line(dot-dash line) and Gluon-Gluon fusion-dash line(dot-dot dash line);

{\bf figure 4}:Total cross section for the process $p \ p \to
Z H_{1}^{0}$ as a function of $m_{H_{1}^{0}}$ for a
$v_{\chi}=1.0(1.5)$ TeV at $\sqrt{s} = 14$ TeV for Drell-Yan-
solid line(dot-dash line) and Gluon-Gluon fusion-dash line(dot-dot dash line);

{\bf figure 5}: BRs for the $H_{1}^{0}$ decays as functions of $m_{H_{1}^{0}}$ for $v_{\chi} = 1.0$ TeV.

{\bf figure 6}: BRs for the $H_{1}^{0}$ decays as functions of $m_{H_{1}^{0}}$ for $v_\chi = 1.5$ TeV.

{\bf figure 7}:Total cross section for the process $p \ p \to
Z H_{1}^{0}$ as a function of $m_{H_{2}^{0}}$ for a
$v_{\chi}=1.0(1.5)$ TeV at $\sqrt{s} = 8$ TeV for Drell-Yan-
solid line(dot-dash line) and Gluon-Gluon fusion-dash line(dot-dot dash line);

{\bf figure 8}:Total cross section for the process $p \ p \to
Z H_{1}^{0}$ as a function of $m_{H_{2}^{0}}$ for a
$v_{\chi}=1.0(1.5)$ TeV at $\sqrt{s} = 14$ TeV for Drell-Yan-
solid line(dot-dash line) and Gluon-Gluon fusion-dash line(dot-dot dash line);

{\bf figure 9}: BRs for the $H_{2}^{0}$ decays as functions of $m_{H_{2}^{0}}$ for $v_{\chi} = 1.0$ TeV.

{\bf figure 10}: BRs for the $H_{2}^{0}$ decays as functions of $m_{H_{2}^{0}}$ for $v_\chi = 1.5$ TeV.

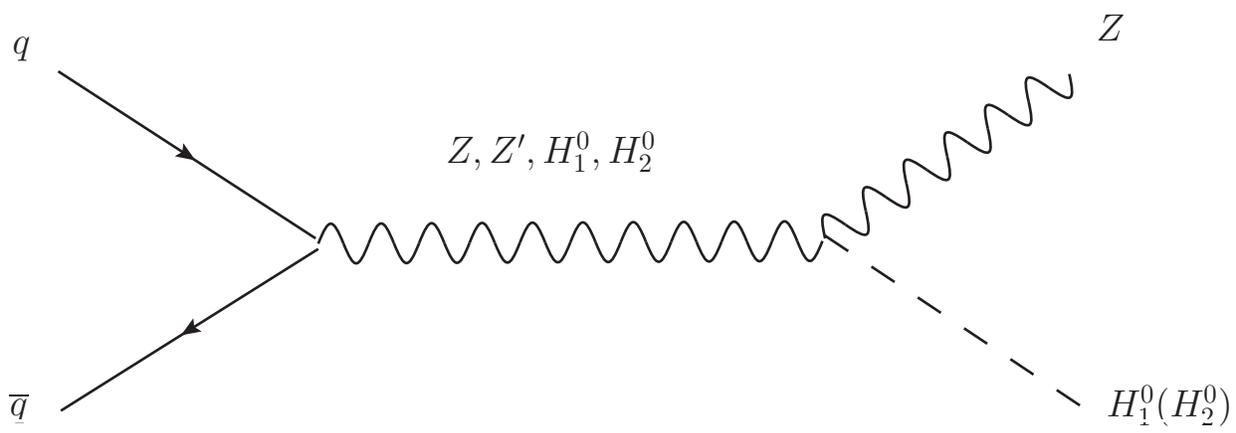
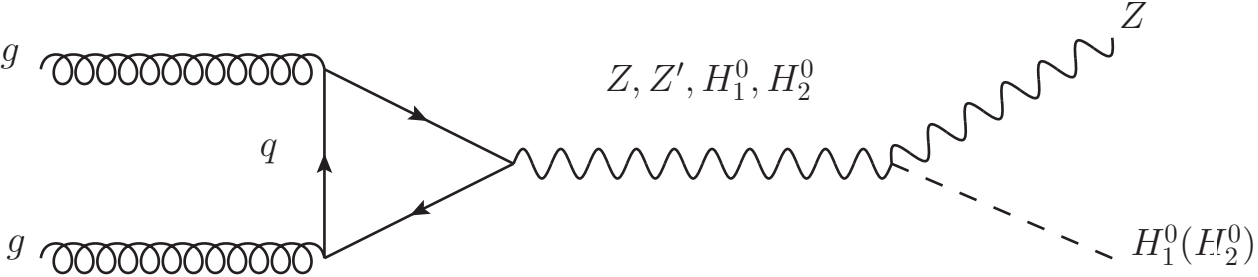
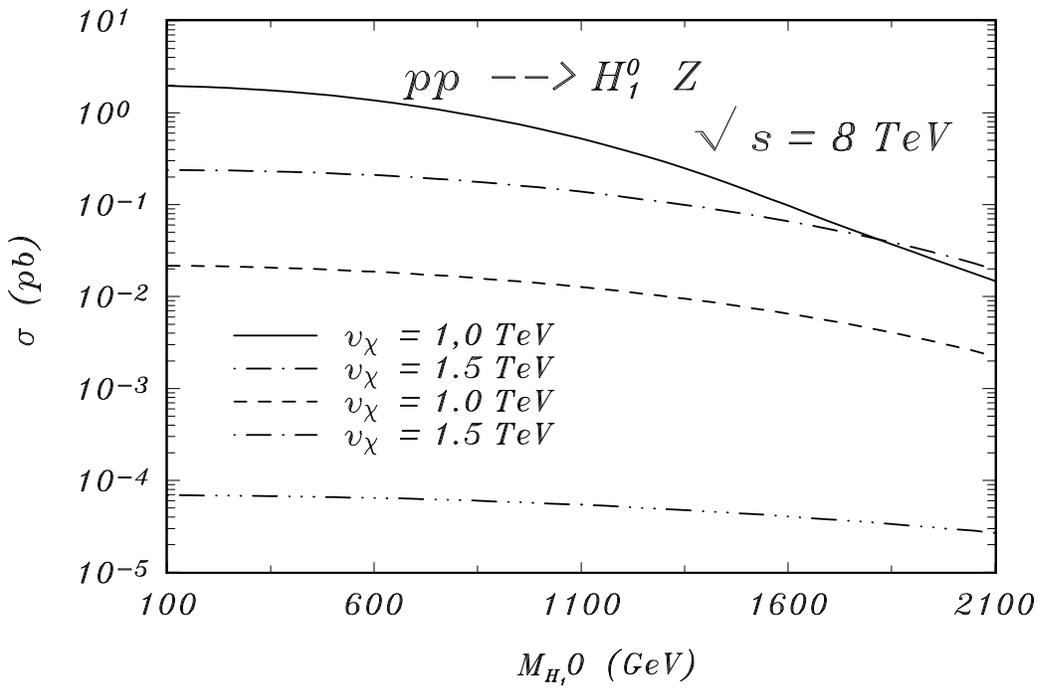
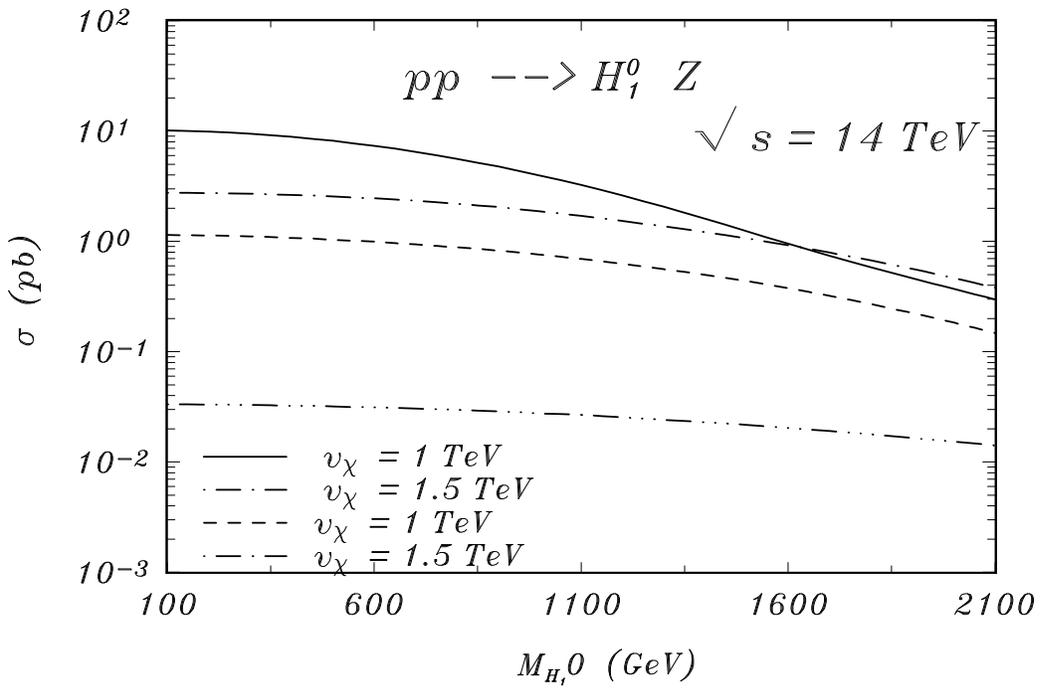
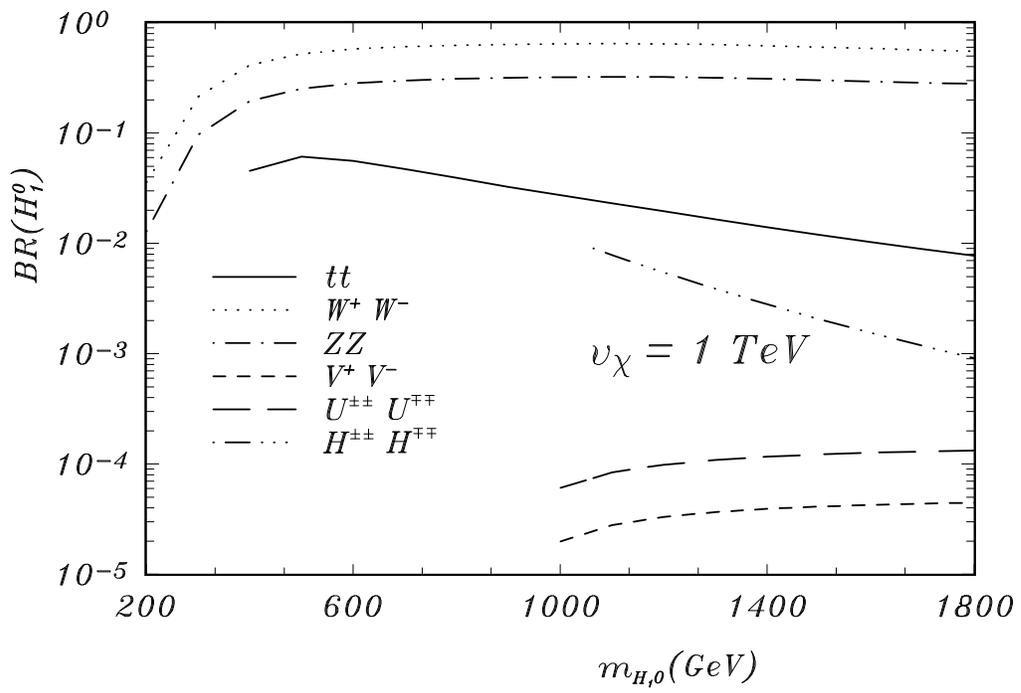
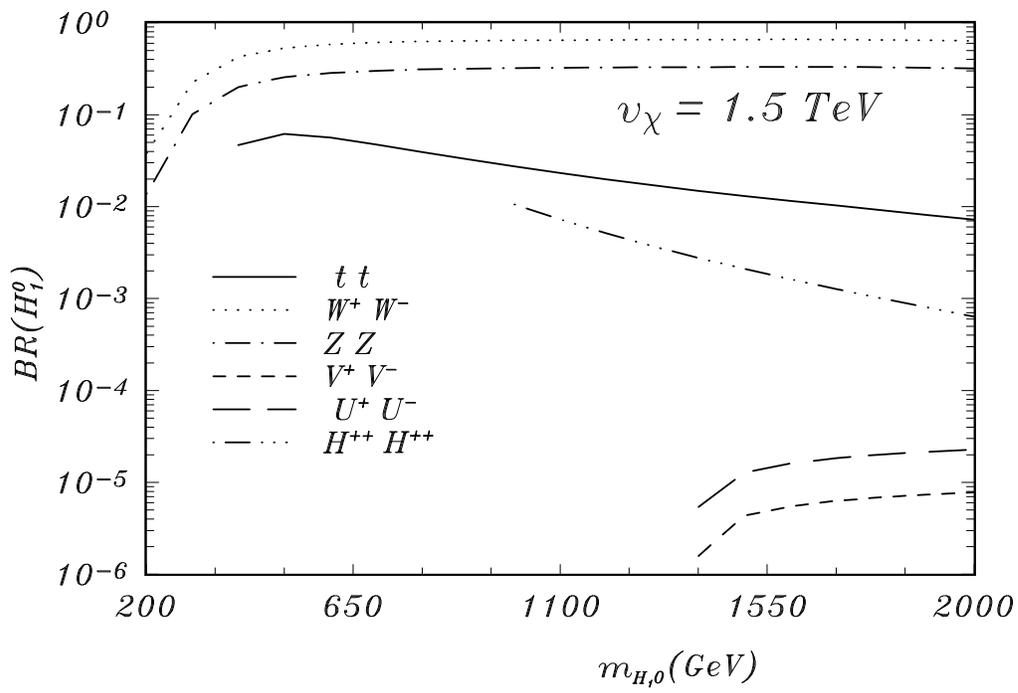
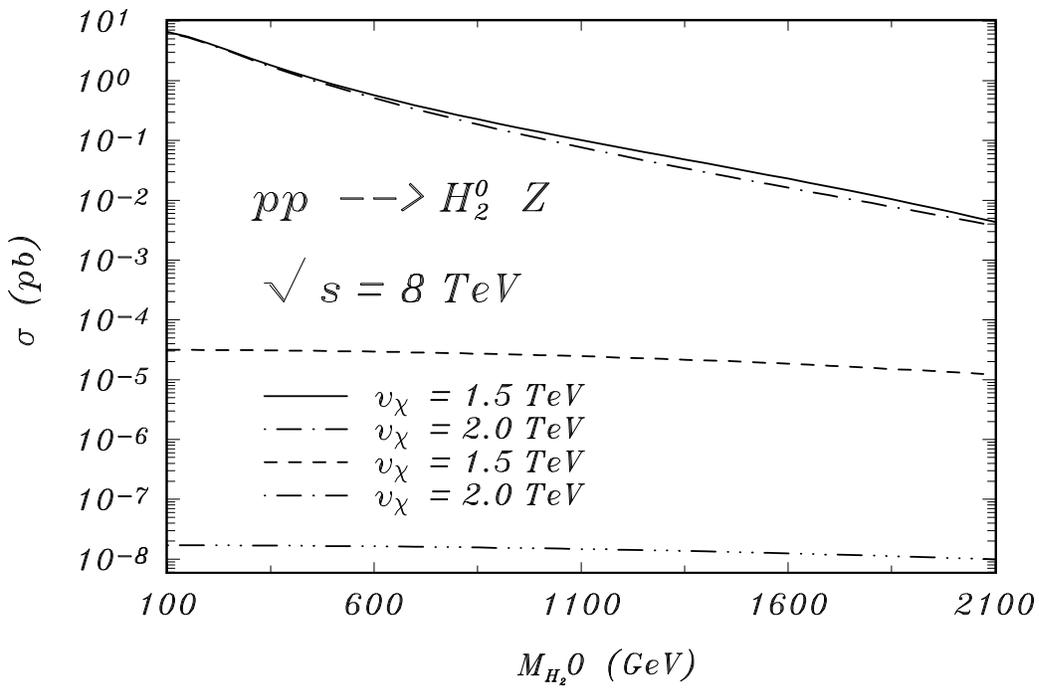
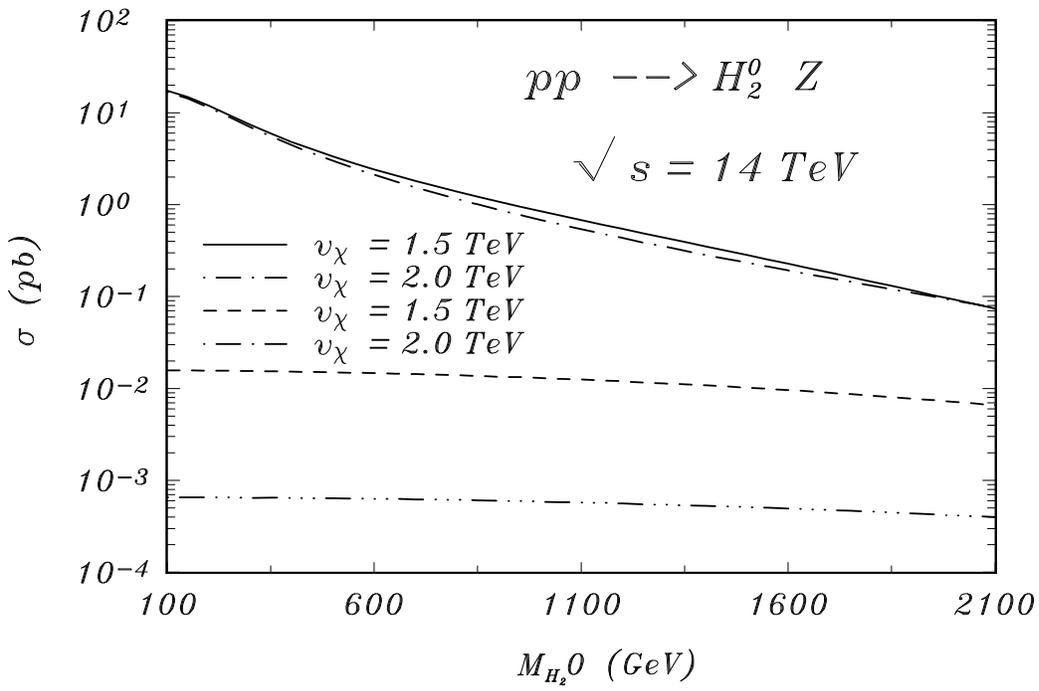
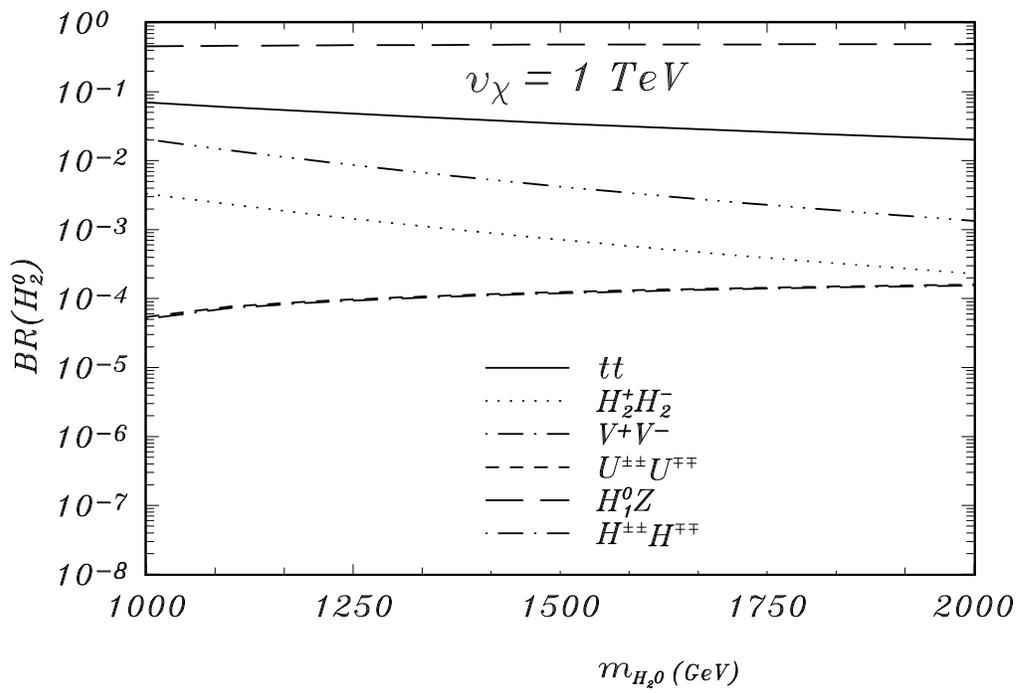
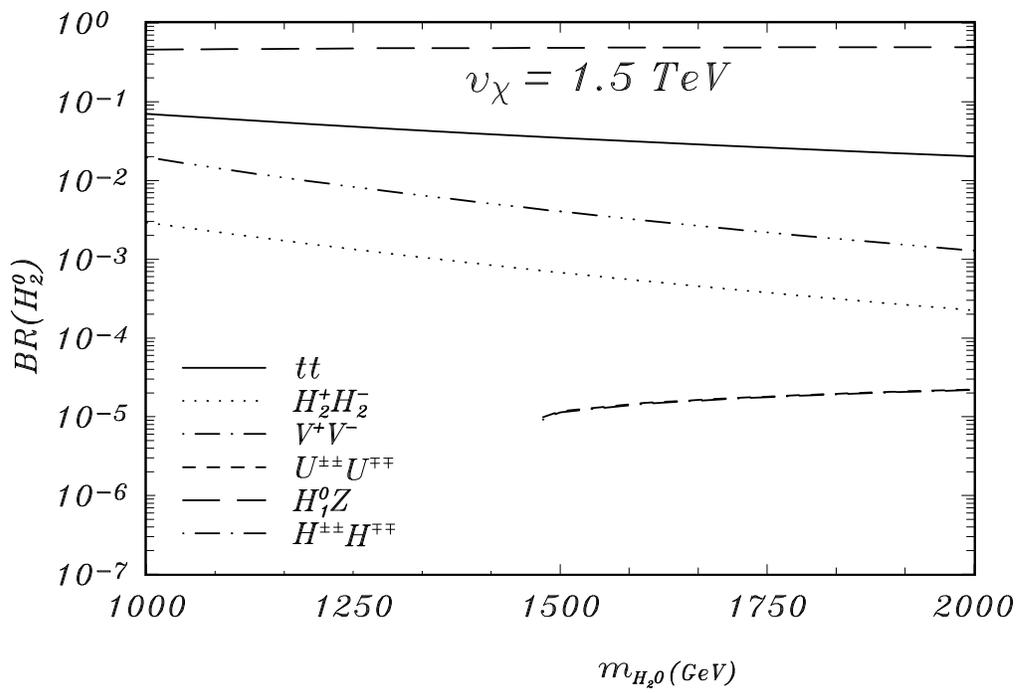
\end{document}